\documentclass[manuscript]{acmart}

\setcopyright{none}

\usepackage{graphicx}
\usepackage{subcaption}
\usepackage{tabularx}

\usepackage[normalem]{ulem}
\usepackage{xcolor}

\usepackage{lipsum}

\usepackage{enumitem}
\renewenvironment{quote}
  {\list{}{\leftmargin=2em \rightmargin=0em}\item\relax}
  {\endlist}
  
\newcommand{\major}[1]{\textcolor{black}{#1}}

\makeatletter
\authorsaddresses{%
  \@mkauthorsaddresses\\[3pt]%
  \textit{This manuscript has been accepted for publication in the journal Behaviour \& Information Technology (BIT)}%
}
\makeatother

\begin{document}


\title[From Search Agents to Dissemination Interfaces]{From Search Agents to Dissemination Interfaces: Understanding Human Trust in Health Information from Conversational Search}



\author{Xin Sun}
\affiliation{
  \institution{University of Amsterdam}
  \country{The Netherlands}
  \institution{and National Institute of Informatics (NII)}
  \country{Japan}
}

\author{Rongjun Ma}
\affiliation{
  \institution{Aalto University}
  \country{Finland}
}

\author{Xiaochang Zhao}
\affiliation{
  \institution{University of Amsterdam}
  \city{Amsterdam}
  \country{The Netherlands}
}

\author{Janne Lindqvist}
\affiliation{
  \institution{Aalto University}
  \country{Finland}
}

\author{Jan de Wit}
\affiliation{
  \institution{Tilburg University}
  \country{The Netherlands}
}

\author{Zhuying Li}
\affiliation{
  \institution{Southeast University}
  \country{China}
}

\author{Abdallah El Ali}
\affiliation{
  \institution{Centrum Wiskunde \& Informatica (CWI) and Utrecht University}
  \country{The Netherlands}
}

\author{Jos A. Bosch}
\affiliation{
  \institution{University of Amsterdam}
  \country{The Netherlands}
}



\begin{abstract}
Large Language Models (LLMs) deployed through Conversational User Interfaces (CUIs) are transforming health information-seeking by offering immediate, interactive experiences compared to traditional search engines like Google. However, how trust is influenced by both the types of search agents and the interface used to disseminate the information remains underexplored. This research integrates two mixed-methods studies (lab sessions and interviews) to comprehensively explore trust perceptions in health information across different search agents and dissemination interfaces. In Study 1 (N=21), we investigated trust in health information sourced from ChatGPT and Google across three types of health-related search tasks. Results showed significantly higher trust in health information from ChatGPT, highlighting the promise of LLM-powered conversational search. Building on this, Study 2 (N=20) extended the investigation to explore how the dissemination interface influences trust in LLM-sourced health information by comparing three interfaces: text-based, speech-based, and embodied, all sourcing from the same LLM. Findings revealed significant trust variations across the dissemination interfaces. Interviews from both studies revealed key factors influencing trust in LLM-powered conversational search, including source credibility, participants' search autonomy, and prior knowledge as well as the interaction style and modality. Our findings highlight the potential of LLM-powered conversational search to transform health information-seeking, underscoring the interplay between the credible search agents and the thoughtfully designed dissemination interfaces in shaping trust. These insights are crucial for developing effective, trustworthy LLM-powered health tools to enhance the health information-seeking experience.
\end{abstract}

\begin{teaserfigure}
  \includegraphics[width=\textwidth]{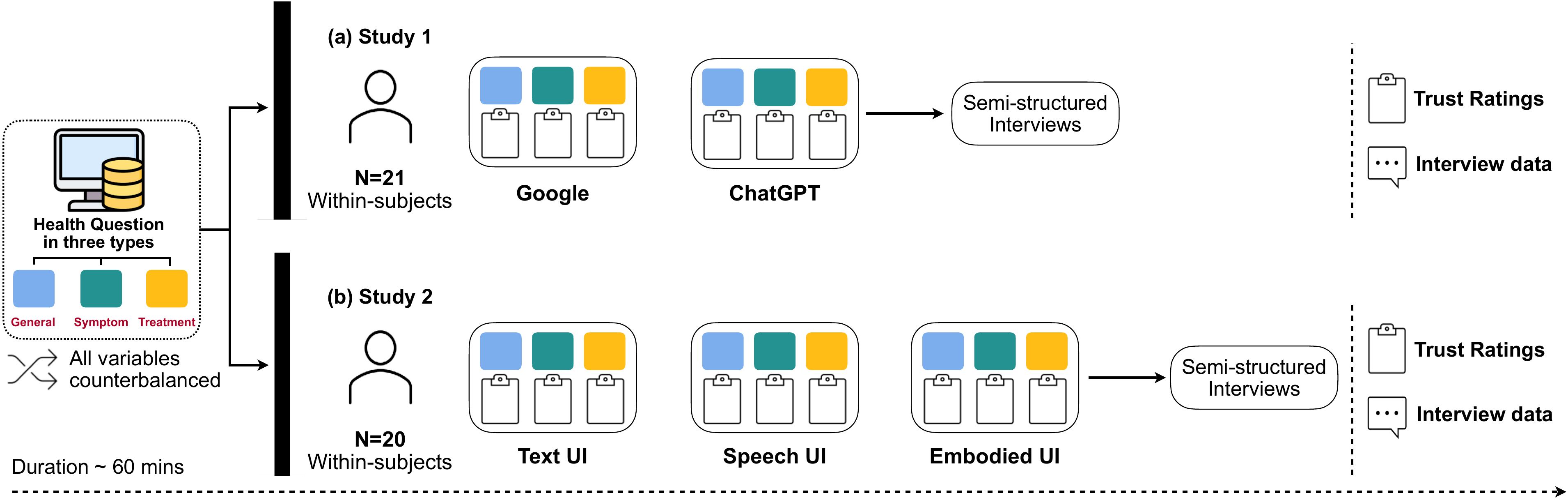}
  \caption{Visual summary of the studies in this paper. (a) Study 1: Lab study to explore the trust perception in health information using two search agents: Google vs. ChatGPT; (b) Study 2: Lab study to investigate perceived trust in health information disseminated by three different conversational user interfaces: text-based, speech-based, and embodied interface.}
  \Description{Figure 1: Visual summary of the study design of the two studies in this paper.}
  \label{fig:procedure}
\end{teaserfigure}



\keywords{Conversational search, User interface, Large language model, Healthcare, Human trust perception}


\maketitle


\section{Introduction}

The reliance on online sources for health-related information has grown significantly due to their convenience and rapid access. Historically, search engines like Google have been the main tools for individuals seeking health information on diseases, treatments, and general wellness advice such as diet and exercise~\cite{Cline, Kibirige, mayo, SILLENCE20071853}. However, the emergence of generative AI has introduced Large Language Model (LLM)-powered conversational agents like ChatGPT as increasingly popular alternatives~\cite{verifiable_source_trust}. 
Unlike traditional search engines that present a list of links to external sources, these LLM-powered agents engage users in interactive dialogues, transforming information-seeking \major{from information retrieval to information generation}, a more personalized and conversational experience~\cite{NEURIPS2020_1457c0d6}. This evolution is particularly significant in health information contexts, where the ease of access, accuracy, presentation and trustworthiness of information directly affect health decisions and outcomes~\cite{t6, t7, t8, hai_1}. 
While extensive research has investigated trust in traditional search engines like Google, the factors influencing trust in LLM-powered conversational search remain underexplored. With the rise of LLMs, people are no longer just searching for web-listed information, they are engaging in dynamic conversations, and the factors shaping trust are now more dynamic, complex, and overall multifaceted~\cite{Vega, Sillence1, chin2002patients}. 
These advancements in search technologies highlight the need for renewed perspectives on how these search agents reshape people's trust in seeking health information online.

While trust in traditional search engines is well-explored, the factors influencing trust in LLM-powered conversational search remain underexplored. 
With the rise of LLMs, conversational search can be conducted through various conversational user interfaces (UIs) ranging from text-based chatbots to voice assistants and embodied avatars~\cite{trust_ui_1, survey_ca_2}.
\major{These CUIs act as essential links between users and health resources~\cite{survey_ca_2}, navigating dynamic and synthesized conversations where the boundary between the search agent and the source is often blurred~\cite{Vega, Sillence1}, and likewise shape user trust~\cite{trust_ui_1,trust_ui_2}. 
Drawing on the MAIN model of trust in technology~\cite{Sundar2007TheMM}, users’ processing and trust of information can vary depending on the interaction style and information modality, such as text, speech, or embodied interactions~\cite{Rheu, trust_interface}.}
Key factors influencing trust in AI-driven health interactions include the source of information, presentation style, and the design of user interfaces that deliver the content. 
For instance, a trustworthy source presenting health advice in a clear style within an intuitive interface may foster greater confidence in the information provided. 
However, embodied interfaces, which simulate advanced human-like traits, add complexity by enhancing engagement while potentially raising concerns about authenticity, privacy, and the ``uncanny valley'' effect~\cite{wikipediaUncannyValley, Anthropomorphism-1, Anthropomorphism-8}. 

Trust is essential in adopting online health advice, influencing how users evaluate information and decide whether to act on it~\cite{SILLENCE20071853}. 
Despite growing interest in LLMs for health information seeking, \major{challenges remain unresolved. 
Specifically, it is unclear how users interactively allocate trust between the health information itself, the search agent providing it, and the LLM-powered conversational user interface through which it is disseminated and presented, and whether these trust mechanisms differ between traditional web-based search engines and LLM-powered conversational search systems. Addressing these gaps is critical for understanding how emerging LLM technologies influence trust in online health information seeking.
}

\major{
Given the foregoing, this work focuses on two perspectives: search agents and dissemination interfaces, and seeks to address the following three research questions:}
\vspace{-2.0mm}
\begin{itemize}
    \item (RQ1) 
    \major{
    Do people’s perceived trust in health-related information differ 
    across (a) search agents (i.e., traditional search engine Google vs. LLM-powered conversational agent ChatGPT)? 
    (b) dissemination interfaces (i.e., text-based, speech-based, or embodied)?
    }
   
    \item (RQ2) 
    Does trust in retrieved online health information correlate with the inherent trust in specific (a) search agents and (b) dissemination interfaces?
    
    \item (RQ3) 
    Which factors contribute to trust in health information across different: (a) search agents and (b) dissemination interfaces?
\end{itemize}
\vspace{-1.0mm}

To address these research questions, we adopted a mixed-method approach, conducting two \major{complementary studies aimed at understanding trust in health information from the perspectives of \textit{search agents} and \textit{dissemination interfaces}.}
In Study 1, we conducted a within-subject lab study ($N=21$) in which participants completed health-related information search tasks using two search agents: Google and ChatGPT. 
Follow-up interviews were conducted to explore the factors influencing participants' trust perceptions.
\major{Building on the insights from Study 1 regarding the potential of LLM-powered conversational search, Study 2 narrowed the focus to isolate the impact of dissemination interfaces.} 
Follow-up interviews were also conducted to further explore participants' motivations behind their trust perceptions.

\major{Our investigation reveals a complex trust landscape shaped by the interplay of agent capabilities and interface modalities.}
First, we observed a distinct \textbf{preference for conversational search}, where the conversational fluency and personalization of LLMs significantly enhanced trust compared to Google's traditional list-based presentation. However, this preference did not extend to multi-modal CUIs. Regarding dissemination, participants favored simple text-based CUIs over speech or embodied ones, citing familiarity level and ease of use as primary drivers.
Second, \major{we identified a critical ``modality-context mismatch''. While human-like textual responses (in Study 1) fostered trust, the explicit anthropomorphism of speech and embodied CUIs (in Study 2), enhanced engagement but triggered privacy concerns and skepticism of authenticity, which negatively influenced the trust. 
This suggests that in high-sensitive health contexts, advancing human-like features does not linearly enhance trust, highlighting the need to balance conversational engagement with perceived safety and usability.}


\major{Our work makes several theoretical and practical contributions.
First, to our knowledge, this work is the first to empirically compare health information seeking between a traditional \textbf{``retrieval''} mode (web search engine) and an emerging \textbf{``generative''} mode (LLM-powered agent), considering both source agents and dissemination interfaces. This helps better understand the role of design, such as usability, familiarity, and interaction style, in shaping trust in retrieved and disseminated health information. 
Second, this study extends beyond the text-dominant focus of prior work by examining multiple dissemination modalities of LLM-powered conversational user interface. We examined existing theoretical models (such as the MAIN model~\cite{Sundar2007TheMM}) in high-stakes health contexts, demonstrating how the interaction and modality influence trust.
Lastly, we offer practical design implications for developing trustworthy eHealth tools that balance engagement with the need for reliability and transparency.}
This lays the foundation for future research to refine trust-building strategies in LLM-powered digital health tools, ensuring that such tools are reliable, user-centered, and capable of effectively addressing diverse user needs in the evolving landscape of generative AI.


\section{Related work}

\subsection{Theoretical Foundations of Trust}
\label{rw_trust_thoery}

\major{Trust is a multidimensional psychological construct central to both interpersonal dynamics and human–technology interactions. 
Classical trust theories conceptualize trust as a willingness to accept vulnerability based on positive expectations of another's conduct. 
Mayer, Davis, and Schoorman’s foundational ABI model~\cite{organizational_trust_model} posits that this willingness is driven by three key dimensions: 
\textit{ability} (the perceived competence of the trustee), \textit{benevolence} (the belief that the trustee acts in the trustor's best interest), and \textit{integrity} (the adherence to acceptable principles). 
In the digital age, Lee and See~\cite{trust_automation} extend these dimensions to automation, defining trust as the attitude that an agent will help achieve the user’s goals despite uncertainty and complexity. 
This perspective is further nuanced by McKnight et al.~\cite{trust_technology_1}, who distinguish between trusting beliefs regarding a technology's attributes and the subsequent \major{trusting intention to depend on that system, such as perceived usability, credibility, and risk.}
}

\major{Understanding and measuring people's trust in online information is a tricky task since human perception of trust is complicated to define and measure. 
Human intrinsic trust is not one-dimensional but is influenced by many factors. 
Past research \cite{20,Gao1,Liu1,SILLENCE20071853,Sillence3} has investigated how people come to trust in mind, providing valuable insights into the way trust is built in online information. 
A critical theoretical distinction for this work lies between \textit{trust} and \textit{credibility}. 
Credibility primarily concerns the believability of information itself, encompassing factors such as accuracy, expertise, and reliability, whereas trust involves the reliance on the agent and the acceptance of vulnerability and risk~\cite{5, 9}. 
Thus, credibility functions as a precursor to trust in information (information-level), while trust in the agent reflects judgments about the provider or system behind the information (agent-level).
Research by Johnson et al.~\cite{ti} identifies that in online health contexts, this formation is driven by the credibility of the source, the reliability and relevance of the content, and the design and usability of the system delivering the information. These factors align with the Technology Acceptance Model (TAM)~\cite{tam3}, where perceived usefulness and usability directly influence user attitudes and acceptance of technology and systems. Further work from Billanes et al.~\cite{trust_tam} validates that trust in technology can influence users' perceived usability, thereby shaping user attitudes and acceptance of technology and systems.
}

\major{
To unify these strands under a coherent theoretical lens, we draw on the MATCH framework~\cite{responsible_ai}, a contemporary model of trust in human–AI communication. MATCH synthesizes three complementary dimensions of trust: 
\textit{Model Attributes}, which refer to the system’s actual competence and reliability (e.g., the quality of retrieved health information); 
\textit{Afforded Cues}, such as interface design and transparency signals that shape user impressions; and 
\textit{Trust Heuristics}, the mental shortcuts users apply under uncertainty, such as relying on brand reputation or the fluency of a response. 
To specifically address the interaction modalities of our study, we complement the MATCH framework with the MAIN model (i.e., Modality, Agency, Interactivity, Navigability)~\cite{Sundar2007TheMM}. 
MAIN specifically explains how specific interface cues, interface modality, alter trust perceptions, particularly in comparing text-based and multi-modal interfaces. 
}

\major{Overall, these trust frameworks (MATCH and MAIN models) provide a robust conceptual backbone for understanding how information quality, system attributes, and user interface cues collectively influence trust formation in search and conversational systems. 
\major{These theoretical foundations clarify the conceptual focus of our work: how users form trust in the information they receive, the agents that provide it, and the interfaces through which it is delivered.}
}

\subsection{Trust in Information and Search Agents for Health Information Seeking}

\major{Online health information searching has become a main way for people to get health advice due to its convenience \cite{Cline, Choudhury, Kim}.
Reflecting the theoretical concepts of trust established in Sec~\ref{rw_trust_thoery}, trust in health information seeking context operates on two distinct levels. 
\textit{Information-level trust} pertains to the user's judgment of the content's intrinsic quality, mapping to the concept of credibility (accuracy, relevance), while \textit{Agent-level trust} concerns the user's willingness to depend on the intermediary system itself. 
This latter form of trust is heavily influenced by the agent's perceived \textit{ability} and \textit{benevolence} (as defined in the ABI model~\cite{organizational_trust_model}) to retrieve or synthesize the information.
}

\major{For decades, web-based search engines like Google have served as the dominant agents for health advice~\cite{Cline, tt, Kibirige}. Search engines, with their powerful ``retrieval'' capabilities, can offer users a wide range of health information.
In this paradigm, the agent functions primarily as a gatekeeper using specific \textit{afforded cues}~\cite{responsible_ai}, such as ranked lists and source snippets, to signal relevance. 
Users typically rely on \textit{trust heuristics} related to the search engine's ranking algorithms to surface reputable sources like medical institutions and websites (e.g., NIH, Mayo Clinic)~\cite{nih,mayo}, while filtering out low-quality commercial content.  
Thus, trust in health information obtained through web-based search is shaped by multiple factors. Many prior works emphasize the role of source credibility, showing that information from well-known medical institutions, or peer-reviewed journals is generally perceived as more trustworthy~\cite{Zhang1,Vega,Sillence1,Liu1,Singal,bates2006effect,dutta2003trusted,hesse2005trust,Lucassen,8}, whereas user-generated or commercially motivated content is often met with skepticism~\cite{5,6,7,8,Choudhury}. However, concerns about the potential for search engine manipulation have arisen~\cite{t5,Seckler,Haque}.
As results, trust in web-based health information seeking is often distributed between information and its provider. While search engines primarily provide links for navigation and ranking, trust perceptions are delegated to external websites that host information~\cite{Lucassen,bates2006effect}. 
This separation allows users to mitigate the potential risk of search engines by independently evaluating the credibility of source websites, particularly when they navigate beyond the search results page.}

\major{Recently, advanced LLMs have entered the health domain, with these agents such as ChatGPT demonstrating increasing capability in responding to health-related queries~\cite{t7,Dalton,t8,t6,verifiable_source_trust}.
These LLM-based agents fundamentally alter trust formation by shifting the interaction from retrieval to generation~\cite{gg_gpt_1}. 
According to MATCH framework~\cite{responsible_ai}, this shift changes the \textit{model attributes} users must evaluate. 
Unlike web search, conversational agents synthesize answers directly, often obscuring the original source. This forces users to rely heavily on the agent's internal \textit{competence} (ability) and \textit{benevolence}~\cite{organizational_trust_model}, as the opportunity for external verification is reduced. 
While studies indicate that LLMs can exhibit high performance in medical queries~\cite{39, t10}, the conversational modality introduces new risks. The \textit{fluency heuristic}, where users conflate grammatical smoothness with factual accuracy—may lead to overtrust, masking hallucinations or biases and lack of verifiable sourcing and transparency~\cite{t6, t9}. Thus, the transparency mechanisms that support trust in Google are replaced by social cues and "personality" in LLMs~\cite{t12, t14}, creating a pressing need to investigate how these differing \textit{afforded cues} influence lay users' trust allocation in high-stakes health contexts.}

\major{Taken together, prior work indicates that trust in health information seeking involves both trust in information content and trust in the search agent, and these trust judgments may operate differently for traditional web-based and LLM-based search agents. Because they differ fundamentally in how they retrieve, generate, and present health information, it remains unclear whether trust established for web-based search extends to LLM-powered conversational search. 
This distinction directly informs our research questions, which investigate how users allocate trust between information and agent, and whether these trust allocations differ when interacting with a web-based search agent (Google) versus an LLM-based conversational agent (ChatGPT) in health information seeking tasks.}

\subsection{Trust in LLM-powered Conversational Search and User Interfaces for Health Information Seeking}


Unlike the extensive literature on trust in web-based search, trust in LLM-powered conversational search remains underexplored. 
Prior work on conversational agents suggests that trust can be shaped by factors such as communication style, perceived medical expertise~\cite{gg_gpt_1}, agent ``personality''~\cite{t14}, consistency of their responses~\cite{t12} and transparency of their generation processes~\cite{t13}. 
Besides, concerns are raised regarding hallucinations and biased content, particularly in high-stakes health contexts~\cite{t6,t9}. 


\major{Crucially, conversational search enabled by LLMs is almost always experienced through conversational user interface (CUIs). It is therefore important to distinguish conversational search as an information-seeking paradigm from CUIs. 
Modern LLM-powered systems operate across both dimensions: they generate health information while simultaneously shaping how that information is perceived through interface cues such as presentation of the synthesized responses and information modality~\cite{CUI_CHI}. 
As a result, trust in LLM-powered conversational search is shaped not only by evaluations of the information itself, but also by users’ perceptions of the conversational UIs that disseminate and present it.
While LLMs can be conceptualized as search agents alongside traditional web-based systems, we focus specifically in this section on how trust in LLM-powered health information seeking is further shaped by CUIs during conversational search interaction, rather than by agent-level attributes alone.}

Prior research shows that trust in CUIs for health information seeking is influenced by multiple interface-level cues. 
Interface design, including information presentation and accessibility, and usability~\cite{fruhling2006influence} play critical roles in fostering trust. 
Persuasive design principles such as social presence and perceived expertise can also enhance trust~\cite{Torning}. 
Luger et al.~\cite{Luger} emphasize the importance of CUIs exhibiting intelligent behavior and contextual understanding to build user trust. Personalizing interactions to individual preferences can increase trust by enhancing perceived system intelligence. 
Consistency with users’ expectations~\cite{Guo2, Luger} and prior experiences~\cite{Luger} also reinforces trust, as does the credibility of information sources~\cite{bates2006effect, Vega, Zhang1, Sillence1}.
\major{Recently, multi-modal CUIs that utilize text, voice, and visual cues to deliver information present unique challenges for trust formation in LLM-powered conversational search. 
Drawing on the MAIN model~\cite{Sundar2007TheMM}, information modality, such as multimodal displays, can serve as ``heuristic cues'' that shape users’ trust perceptions~\cite{Sundar2007TheMM}.
While these multi-modal interactions can enhance user engagement and understanding~\cite{multimodal_ca_engagement}, maintaining consistent reliability across modalities is crucial.  
Moreover, characteristics of voice and visual representation can also influence trust~\cite{multimodal_trust_factors_1, voice-1, voice-3, embodied_trustworthy_design_ethics}. 
Therefore, providing consistent and reliable information and interaction across modalities is essential to sustain trust.
Applied to LLM-powered conversational search, this perspective suggests that trust judgments may differ depending on how health information is generated by LLM-based search agents and how it is disseminated and presented through CUIs.
}


\major{Despite rapid advancements, existing research on LLMs in health contexts has focused primarily on evaluating information quality, with limited attention to how users form trust during conversational interaction. 
In particular, little is known about how different CUIs influence trust in LLM-generated health information and how interface-level cues interact with the trust of information. 
Addressing this gap motivates our research questions and empirical investigation on trust formation in LLM-powered conversational search across different multi-modal conversational user interfaces.
}



\section{Study 1: Comparison of search agents for health information seeking}
We conducted a mixed-methods study involving a laboratory-based study followed by semi-structured interviews. In the lab study, we employed a within-subjects design to examine how participants interacted with outputs of two different search agents: the search engine Google and the LLM-powered agent ChatGPT (i.e., GPT-4o). 


\subsection{Study Methods}


\subsubsection{\textbf{Participants}}

A power analysis using G*Power \cite{gpower} indicated that a minimum of 20 participants were needed to detect a medium effect size of 0.25 with an alpha level of .05 and 80\% power. Accordingly, we recruited 21 participants (N=21) through the institute's 
recruitment system. Participation was voluntary and informed consent was obtained before the lab sessions. 
Each participant received a 
monetary compensation for their time according to the institute's guidelines. Participants were required to be proficient in English and experienced with online information searching. 
The study received approval from the ethics and data protection committee of the institute (ID: XXX).

The characteristics of participants are summarized in Table \ref{table:demographic_search}.
Most participants (71.4\%) were between 18 and 24 years old. Regarding education, 47.0\% have undergraduate degrees, 47.6\% had a postgraduate qualifications, and 4.7\% held doctorate degrees.
Participants represented various professional fields: 47.6\% from social sciences, 19.0\% from business and commerce, 14.3\% from computer and information technology, and 19.1\% from other sectors. 
In terms of online health information seeking, 42.8\% reported frequent use of online sources, 52.3\% occasionally, and 4.7\% stated they rarely or never relied on online resources.

\begin{table}[!h]
\centering
\footnotesize
\renewcommand{\arraystretch}{0.90}
\begin{tabularx}{\columnwidth}{p{5cm} >{\raggedright\arraybackslash}p{6cm} >{\raggedright\arraybackslash}X}
\toprule
\textbf{Demographic} & \textbf{Categories} & \textbf{Numbers of Participants (\%)} \\
\midrule
Gender & & (N=21)
\\ 
 & Female & 16 (76.1\%)
\\ 
 & Male & 5 (23.9\%)
\\
\hline
Age & & 
\\
 & 18-24 & 15 (71.4\%)
\\
 & 25-34 & 5 (23.8\%)
\\
 & 65+ & 1 (4.8\%)
\\
\hline
Education & & 
\\
 & High school & 1 (4.7\%)
\\
 & Bachelor & 9 (42.9\%)
\\
 & Master & 10 (47.6\%)
\\
 & Doctor & 1 (4.7\%)
\\
\hline
Professional Domain & & 
\\ 
 & Social Science & 10 (47.6\%)
\\
 & Business and Commerce & 4 (19.0\%)
\\
 & Health and Medical Science & 1 (4.7\%)
\\
 & Computer Science \& Information Technology & 3 (14.3\%)
\\
 & Other & 3 (14.3\%)
\\
\hline
Frequency of online health \\information seeking & & 
\\
 & Often & 9 (42.8\%)
\\
 & Sometimes & 11 (52.3\%)
\\
 & Rarely & 1 (4.7\%)
\\
\hline
Frequently used search agent & & 
\\
 & Search engine & 21 (100\%)
\\
 & Conversational agents & 15 (71.4\%)
\\
 & Social media platforms & 8 (38.0\%)
\\
\bottomrule
\end{tabularx}
\vspace{0.1mm}
\caption{Characteristics of participants in Study 1.}
\vspace{-8mm}
\label{table:demographic_search}
\end{table}


\subsubsection{\textbf{Search tasks}}
\label{search_tasks}
In this study, search tasks are health-related questions that we ask participants to find answers to during the lab session. 
Each participant completed six search tasks, selected from an open-sourced dataset \cite{CHQ_Summ}, which aggregates personal health questions from Yahoo~\cite{yahoo} labeled by types.
We pre-selected 25 questions from each of the three types to ensure diversity and comprehensiveness. The complete list of questions used in the study is included as~\textbf{Supporting Material}.
Three types of health questions are described as follows:  

\textit{General health questions}.
This group of questions aims to gather general knowledge or facts about a specific health topic. Examples include: \textit{"Do you have information about Weight Control?"}; \textit{"Do you have information about Vitamin D?"}

\textit{Symptom and cause-related health questions}.
This group of questions revolves around understanding symptoms or causes associated with particular health conditions. The example questions are: \textit{"What are the symptoms of the eating disorder?"}; \textit{"What causes Memory loss?"}

\textit{Treatment-related health questions}.
This group of questions seeks information about potential treatments for specific conditions. Examples include: \textit{"What are the treatments for dry eye syndrome?"}; \textit{"How can I lower my heart rate?"}

\major{
We define ``health information'' differently across the search agents given the distinct mechanisms: for search engine Google, information encompasses both the results page snippets and the content of external websites; for LLM-based search agent ChatGPT, it refers exclusively to the direct, synthesized textual responses generated by the LLM.
}


\subsubsection{\textbf{Search agents}}
\begin{figure}[htbp]
\centering
\includegraphics[width=0.97\textwidth]{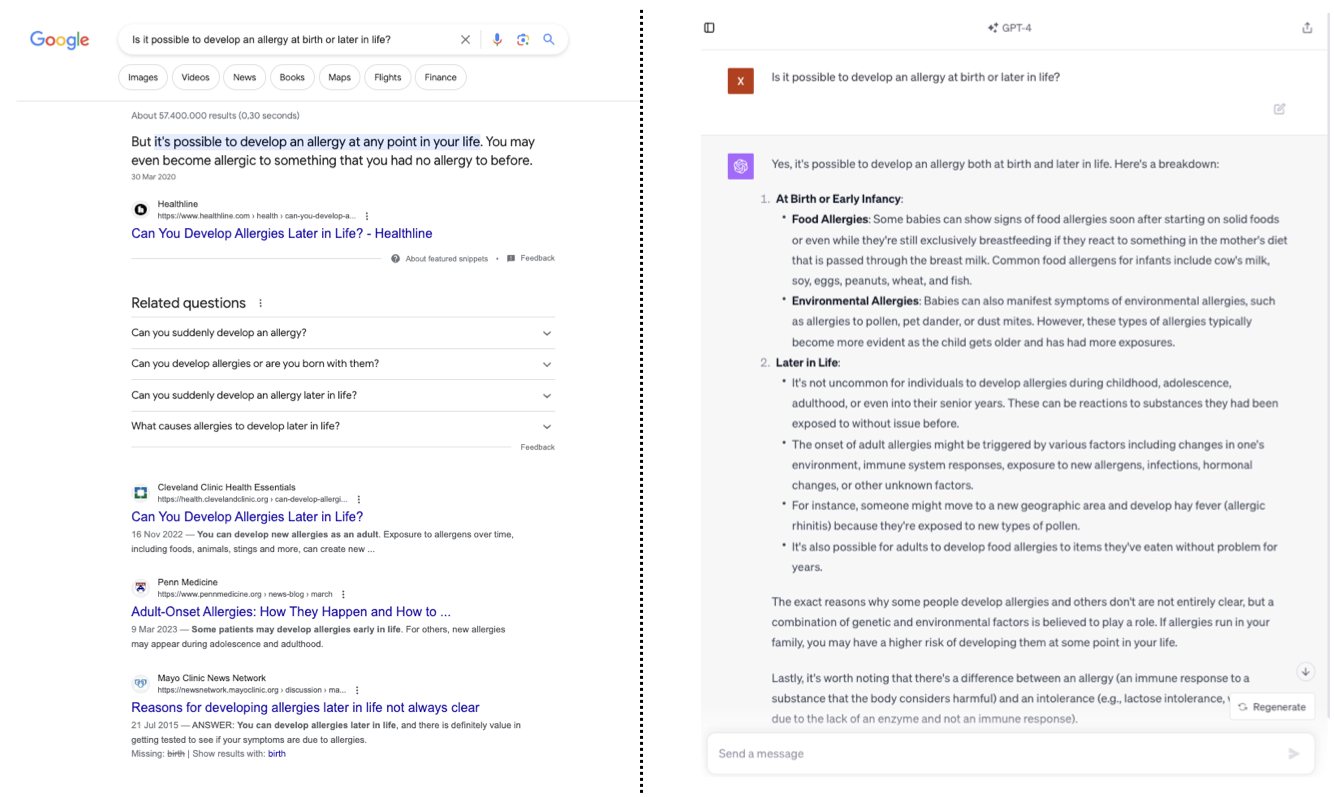}
\caption{Interface of search agents with a search task example using the first type of search task (informational health questions): 'Is it possible to develop an allergy at birth or later in life?'}
\label{fig:search_agent}
\end{figure}

We used two search agents for health information in this study as shown in Fig~\ref{fig:search_agent}: Google and ChatGPT. The study was conducted in the institute lab in a quiet room with desktop computers running the Windows operating system. Participants accessed both search agents via Chrome browser (version 117.0.5938.92) by visiting their official websites. 
For interactions with ChatGPT, participants used the GPT-4 model (without web browsing capability).
This setup ensured all participants had a consistent experience while using up-to-date technology for each search agent.

\vspace{-2mm}


\subsubsection{\textbf{Measures}}
Upon providing consent, we collected demographic information and their usage experience with different search agents. During the study, we measured participants' trust in health-related information obtained from two search agents. Below, we detail the specific measures used.

\textit{Propensity of trust in technology.}
To gauge participants' inherent trust in technology, we employed the questionnaire \cite{ppt} that assessed general tendency to trust technology. This scale consists of six items rated on a 5-point Likert scale (1=Strongly Disagree to 5=Strongly Agree). An example item was: \textit{"I think it’s a good idea to rely on technology for help."}

\textit{Human perceived trust in health-related information.}
After each search task, participants rated their trust in the retrieved health information. We used the validated 'Trust of Online Health Information' questionnaire \cite{ti,Rowley2015StudentsTJ} to assess the credibility, reliability, and believability of information. This survey included eleven items rated on a 5-point Likert scale with a Cronbach's alpha of .95 in our study. An example item was: \textit{"The information appears to be objective."}

\textit{Human perceived trust in the search agent.}
Participants rated their trust in search agents' ability to deliver health information after completing all three search tasks with each agent. The 'Model of Online Trust of Health Information Websites' questionnaire \cite{tt}, consisting of 15 items on a 5-point Likert scale (Cronbach's alpha=.62) was used to assess trust in agents. An example item was: \textit{"The search agent (e.g., Google) provides truthful information."}

\major{
We assessed participants' prospective behavioral intention to use a specific search agent employing a single item adapted from the Technology Acceptance Model (TAM)~\cite{tam3}: \textit{``I would intend to use Google (or ChatGPT) for health information seeking in future.''}
This metric was included to evaluate whether the trust formed during the session translates into potential real-world adoption, consistent with prior findings establishing trust as a key predictor of technology acceptance (cf., Billanes et al.~\cite{trust_tam}).
}


\subsubsection{\textbf{Study procedure}}
The procedure of Study 1 is illustrated in Fig~\ref{fig:procedure}. 
Following informed consent, we employed a within-subjects design where participants completed three search tasks using Google and three using ChatGPT, with the order of agents and tasks counterbalanced to mitigate order effects.
\major{For each task, participants were asked to enter a query, review the results, with the specific requirement of clicking at least one result when using Google. 
Participants were free to reformulate their query and to interact freely with the assigned search agent until they found a satisfying answer.} 
Participants rated their trust in information after each task and their overall trust in the search agent after completing each condition block.

The lab session concluded with a 25-minute semi-structured interview exploring participants' interaction strategies and comparative reasoning of their trust ratings regarding the two search agents. 
The interviews followed a progression from general daily health search habits to specific experiences and validation strategies with each search agent. The interview session concluded with a comparative discussion of two search agents and considerations for potential improvements.
The full interview protocol is provided as the \textbf{Supporting Material}.


\subsubsection{\textbf{Data analysis}}
\label{qualitative_analysis}
We collected both quantitative and qualitative data to understand how people interact with two different search agents for health information and assess trust in the health information obtained from these agents. 

For quantitative analysis, we first verified statistical assumptions using Shapiro-Wilk (normality)~\cite{SHAPIRO1965} and Bartlett's (homogeneity)~\cite{Arsham2011} tests. 
We then examined trust differences using two-way repeated-measures ANOVA~\cite{anova} and paired samples t-tests~\cite{t_test}, complemented by correlation analyses to explore variable relationships.

We conducted a thematic analysis \cite{braun2012thematic} of the qualitative interview data. Initially, the first three authors transcribed the interviews to familiarize themselves with the data. Subsequently, three coders open-coded the same transcript independently to develop an initial codebook that guided further coding. \major{When discrepancies arose, we resolved them through discussion, in line with qualitative approaches that consider multiple perspectives as analytically valuable for refining codes and exploring alternative meanings \cite{barbour2001checklists}. This process ensured shared agreement on how codes were phrased, the level of abstraction applied, and the analytic focus guiding the coding decisions.} Then three coders divided and independently coded the remaining interviews. Codes were iteratively refined and expanded as new insights emerged. 
We did not calculate inter-rater reliability (IRR) \major{because our analytic goal was to interpret the data collaboratively and to develop a shared understanding of the material rather than quantify agreement through numerical metrics~\cite{mcdonald2019reliability}. Instead, we supported credibility and consistency through iterative team based coding, regular reconciliation meetings, and cross checks of coded subsets~\cite{mays2000assessing}.}
Together, they discussed themes relevant to our research questions, identifying patterns that emerged across the interviews.

\subsection{Quantitative Findings: Lab Sessions}


\subsubsection{\textbf{Descriptive statistics}}

We measured participants' general trust in technology, their trust in health information provided by two search agents, and their trust in the agents themselves. The descriptive results are presented in Table \ref{table:descriptive_search}.
The average propensity of trust in technology among participants was moderately high (M=3.62, SD=.76 on a 5-point scale), indicating a general positive trust towards technology. Trust in health information varied between search agents; participants generally trusted information from ChatGPT more than from Google, with mean scores of 4.05 (SD=0.47) for ChatGPT and 3.77 (SD=0.64) for Google. Trust in the search agents also reflected a preference for ChatGPT over Google, with mean scores of 3.58 (SD=.98) for ChatGPT and 3.31 (SD=.99) for Google.

\begin{table}[htbp]
\centering
\renewcommand{\arraystretch}{0.90}
\begin{tabularx}{\columnwidth}{p{5cm} >{\raggedright\arraybackslash}X >{\raggedright\arraybackslash}X}
\toprule
\textbf{} & \textbf{Google \newline Mean (SD)} & \textbf{ChatGPT \newline Mean (SD)} \\
\midrule
Trust (search task 1)  & 3.69 (.90) & 4.01 (.74)
\\ 
Trust (search task 2) & 3.81 (.89) & 4.10 (.74)
\\ 
Trust (search task 3) & 3.82 (.88) & 4.03 (.81)
\\ 
Trust (Avg of three tasks) & 3.77 (.64) & 4.05 (.47)
\\ 
Trust in search agent & 3.31 (.99) & 3.58 (.98)
\\
Intention to use & 3.57 (.49) & 3.52 (.85) 
\\
\bottomrule
\end{tabularx}
\caption{Descriptive statistics of variables in Study 1. Search task 1 is General health questions, search task 2 is Treatment-related questions, and search task 3 is Symptoms and Diagnosis-related questions.}
\vspace{-5mm}
\label{table:descriptive_search}
\end{table}


\subsubsection{\textbf{Human trust in health-related information differs by search agents}}
\major{To address RQ1a, which asks whether perceived trust differs between traditional and LLM-powered search agents,} we conducted a two-way repeated-measures ANOVA to compare differences in trust between two search agents. The analysis was performed to explore if there are differences in trust, including trust in the retrieved information and search agents.  




\begin{figure*}[!ht]
\centering
\vspace{-1.0mm}
\includegraphics[width=0.9996\textwidth]{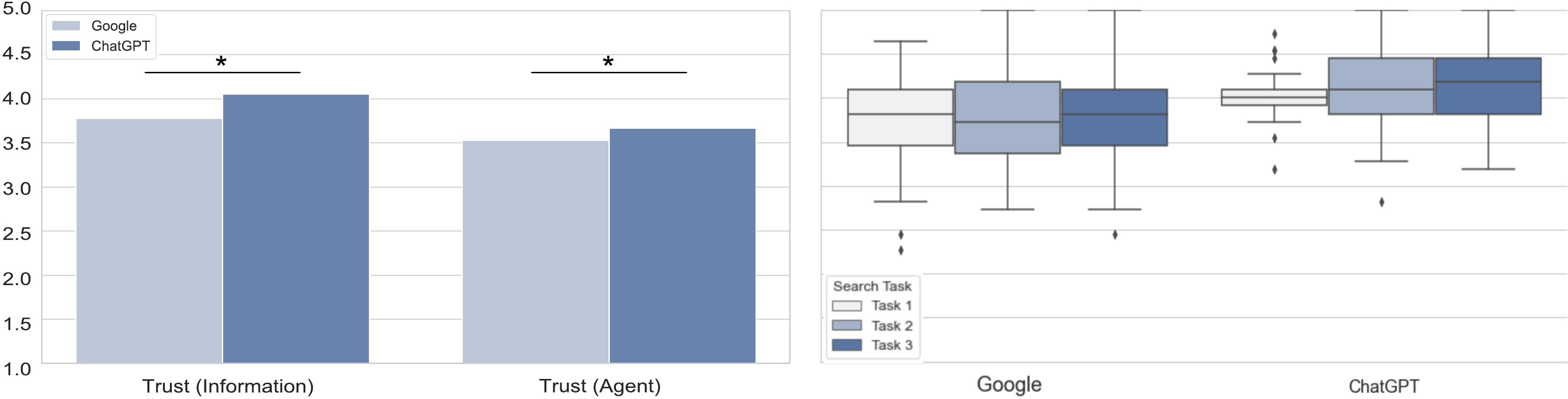}
\vspace{-3.0mm}
\caption{Left: Mean of trust in health information, the search agent, and the intention to use the search agent for health information search. Right: Mean of trust in health information between two search agents across three search tasks. (*$p$<.05)}
\vspace{-2.0mm}
\label{fig:search_mean_difference}
\end{figure*}


As illustrated in Fig~\ref{fig:search_mean_difference}, the analysis revealed a significant main effect of the search agent on trust in health-related information, \( F(1, 20) = 6.73, p = .017 \), and \( \eta^2 = .057 \). This suggests a notable difference in trust, with ChatGPT generally being trusted more than Google for health-related advice. 
Trust levels did not show significant changes across different types of search tasks, \( F(2, 40) = 0.63, p = .480 \), with \( \eta^2 = .006 \). This indicates that the type of health question did not significantly influence trust in the retrieved information.
The interaction between search agents and task types was not significant, \( F(2, 40) = 0.20, p = .777 \), and \( \eta^2 = .002 \). This indicates that trust differences between the agents were consistent across various search tasks, confirming the stability of trust across different types of search tasks.


Additionally, we did a paired sample t-test to explore the human perceived trust in the search agent itself. The trust scores in Google and ChatGPT as search agents are statistically significant with the difference in means as 0.27, $t$(21)=-2.53, $p$=.02, Cohen's $d$=0.55 (see in Fig \ref{fig:search_mean_difference}). 


\subsubsection{\textbf{Correlation of trust across search agents}}

\begin{figure*}[!h]
\centering
\vspace{-2.0mm}
\includegraphics[width=0.88\textwidth]{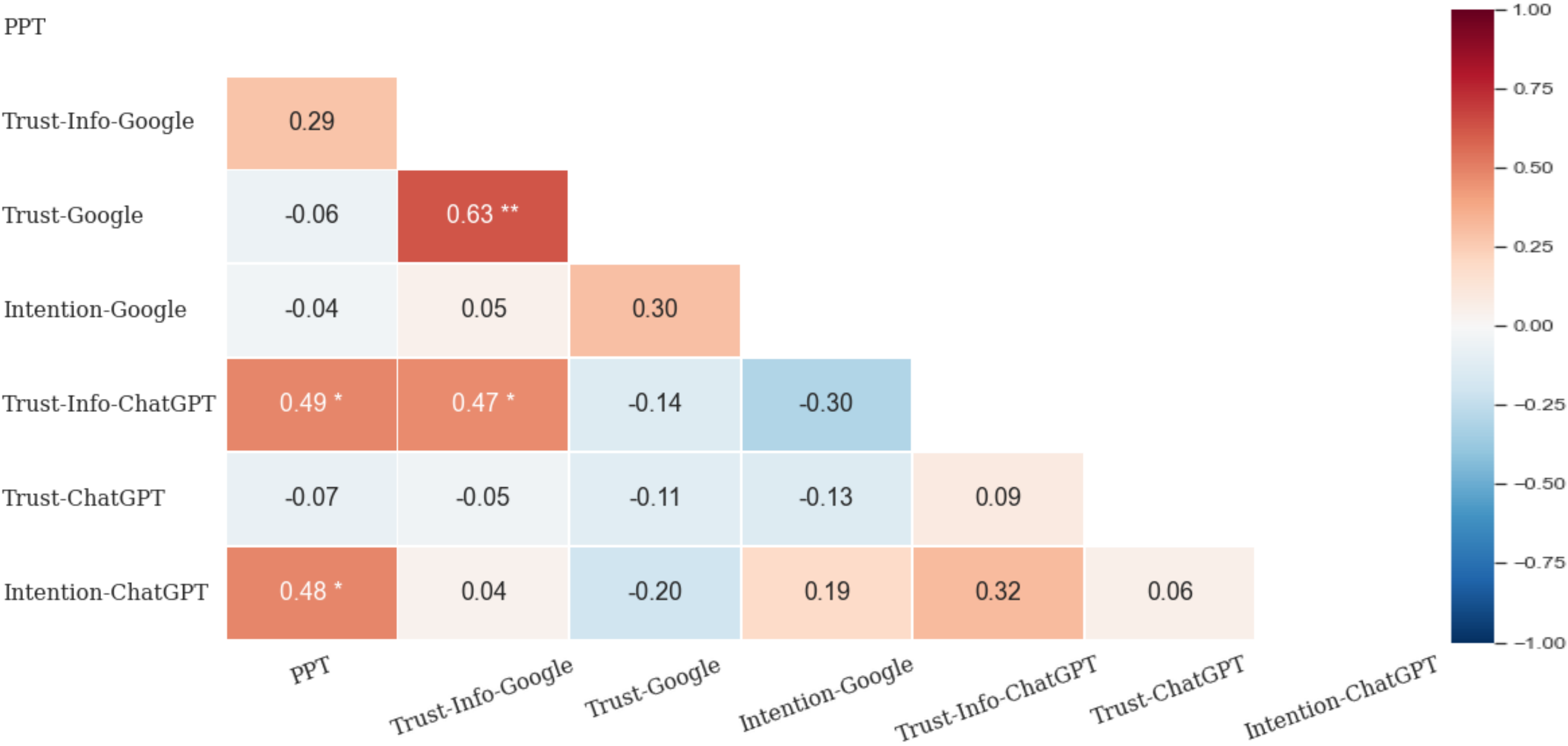}
\vspace{-1.0mm}
\caption{
Pearson correlation with key variables (**$p$<.01, *$p$<.05). Note: 
"PPT" means participants' propensity of trust in technology; 
"Trust-Info-" represents perceived trust level in health information obtained; 
"Trust-" means trust level in the search agent.}
\vspace{-3.0mm}
\label{fig:search_correlation}
\end{figure*}

\major{To investigate RQ2a regarding the relationship between inherent trust in search agent and trust in information,} we conducted Pearson correlation analyses (detailed in Fig~\ref{fig:search_correlation}).

We found a significant positive correlation between trust in information and trust in the search agent for Google \major{($r$(21)=0.63, $p$=.003)}. 
\major{However, this relationship was not significant for ChatGPT ($r(21)=0.09, p=.21$).}
Regarding the propensity of trust in technology, we observed significant positive correlations with both trust in information from ChatGPT ($r(21)=0.49, p=.02$) and the intention to use ChatGPT ($r(21)=0.48, p=.026$), but it does not extend to trust in ChatGPT as a search agent \major{($r$(21)=-0.07, $p$=.16)}.. 
Conversely, this general technology trust showed no significant correlations with Google, either as a source of information ($r$(21)=0.29, $p$=.20) or as a search agent \major{($r$(21)=-0.06, $p$=.73)}.
The correlation analysis not only strengthens our understanding of how different facets of trust are interconnected, but also lays the groundwork for more complex analyses and interpretations in our qualitative findings.

\subsection{Qualitative Findings: Semi-Structured Interviews}

\major{To identify the specific factors that contribute to perceived trust across both search agents (RQ3a),} we report on themes generated from interview data to understand the formation of trust.
The findings are organized into two parts: \textit{How do people search for health-related information} provides a contextual overview of how individuals utilize online sources to find health-related information and offers insights into the scenarios in which trust is shaped. \textit{What factors influence trust in information} illustrates the factors that influence trust in information retrieved from two different agents. 

\subsubsection{\textbf{How do people search for health-related information?}}

\paragraph{\textbf{Searching online is a pre-step to deal with health-related problems.}} Although our participants regularly search online for various health-related topics, none face significant health concerns. The major topics for online searching include\textit{diet and exercise} (P2, P14-15, P17, P19, P21), accident treatment such as \textit{tore a muscle} (P2), personal health conditions such as \textit{allergies} (P6), and \textit{mental health issues} (P4). Since most participants consider health-related information serious, they approach information searching cautiously, withholding a degree of trust because \textit{``Everything I see on the Internet isn't necessarily true''} (P4).
Participants typically don't expect to find a perfect solution when searching online. Instead, they aim to refresh their existing knowledge, learn general information about the topic, or gain direction for the next step. After searching, participants selectively adopt online advice 
depending on the importance and urgency of health matters, as well as the practicality and risks of the advice.
\begin{quote}
    \textit{``So if it says like it take a pill or something like that, I won't take it. But if says like, drink water, stay in bed, of course, I'll do that.'' - P10}
\end{quote}

\paragraph{\textbf{People use their own combinations of sources for health-related information.}} 
People have their own preferences for health information sources, including forums, social media, and certain websites. Participants recognize the credibility and practicality of specific resources based on their experience. For instance, many participants trust official web domains like the NHS website and Google Scholar (P1-2, P4-12, P14-21). Some participants naturally trust big brands like Google (P7). Others consider social media reliable because influencers' demonstrations help them match symptoms more practically (P11, P17, P19).
Additionally, people do not rely on a single resource but a combination of multiple sources. \textit{``Double-check if another source says the same, make a comparison''} (P9) helps participants validate the information.
To incorporate ChatGPT into the landscape of information sources, participants suggest a scenario where ChatGPT is used as a preliminary step before conducting searches on Google:
\begin{quote}
    \textit{``But also if I didn't know anything about a health issue, I could look it up on the ChatGPT first to also like get a sense of direction. What this could be related to and then continue my search on Google.'' - P1}
\end{quote}

\subsubsection{\textbf{What factors influence trust in information?}}
\label{factors_agents}
\paragraph{\textbf{Previous experience of using an agent influences people's trust.}}
Judgment on the trustworthiness of retrieved information largely depends on participants' prior knowledge of the topic, which helps them assess the information. However, several factors differentiating ChatGPT from Google also influence trust.
Previous positive and negative experiences influence people's current trust in search agents. Positive experiences of recovering after following online advice increase trust in that source (P11), whereas negative experiences with misinformation create uncertainty:
\begin{quote}
    \textit{``ChatGPT can give me some results really quickly, but if I copy the DOI number and double-check it, you'll find they're totally wrong. [...] So that's one of the important reasons I don't fully trust ChatGPT.'' - P8}
\end{quote}

\noindent Compared to Google, ChatGPT is a new tool with which most participants have limited experience. While people are still figuring out how to search for health-related information with ChatGPT, their trust may be influenced by \textit{``news and rumors''} (P21). On the other hand, many participants use Google by default, but trust is weakened by experiences like excessive commercial information (P1-2, P4, P10-11, P13, P16, P18-19, P21) and \textit{``filter bubbles''} (P7) that isolate information to personalize searches\cite{Filterbu39:online}.

\paragraph{\textbf{Information presentation influences people's trust.}}
Our participants believe that presenting information in a professional yet understandable form is crucial for trust. This includes language expression, information structure, and visual cues.
First, is the language expression. Our participants feel the medical field is highly specialized and health information often contains terminology. They think that professional use of language and terminology enhances the credibility of the information but also makes it harder for them to understand the topics they're searching for (P1, P2, P17). 
Besides terminology, the tone of language expression also affects trust, such as the use of passive voice and confidence in expression. The passive voice helps convey objectivity, leading to greater trust (P17). Furthermore, the confidence with which ChatGPT delivers information significantly impacts participants' trust. 
An uncertain answer that includes terms like ``\textit{maybe}'' or ``\textit{I'm not sure}'' (P10) lowers trust. However, an overly certain answer can also appear untrustworthy:
\begin{quote}
   \textit{``if it's too, too confident in stating this and this leads to this [...], (I'll) trust less, Because the answer it's never so straightforward.'' - P16} 
\end{quote}

\noindent Other factors, such as information structure and visual cues, also play a role in shaping trust. A well-structured response guides participants to understand the result, and a logical flow enhances the reliability of the information (P11). In the context of Google, visual cues, such as being "very colorful" (P14) or having "unprofessional logos on the websites" (P4), can lead to lower trust regardless of the content of the information.

\paragraph{\textbf{Autonomy influences trust}}
In most cases, control of the search process increases people's certainty and trust in the answer. Google provides more control, giving participants autonomy in each step. Access to rich information enhances trust because ``\textit{with Google you have more the feeling that you have your own judge}''(P13). However, this method is not always preferred, especially when lacking knowledge for evaluating the information.
\begin{quote}
   \textit{``The more information you see, more overwhelming it also get, so you may find out you don't know everything from googling too much. And then ChatGPT typically make it easier and give you (information) more trustworthy.'' - P16}
\end{quote}

\noindent On the other hand, our participants shared that if they have enough domain knowledge to evaluate answers from ChatGPT, a straightforward and highly relevant answer from ChatGPT is preferred (P1, P5, P7, P9-11).

\paragraph{\textbf{Human-like interaction influences trust}}
\label{sec:human-like1}
The major difference between searching with Google and ChatGPT lies in the interaction. The way participants interact with agents influences their trust, and preferences for these interactions vary depending on the situation. 
Many participants compare interacting with ChatGPT to talking with a human. The human-like dialogue with ChatGPT allows participants to follow up on specific details and receive personalized responses. This context-aware process enhances trust, particularly in personal situations, such as when participants are trying to understand potential causes of pain in specific parts of the body (P11).
\begin{quote}
    \textit{``You can just ask it (ChatGPT) like you would ask your friend and it understands what you're talking about.'' - P2}
\end{quote}

\paragraph{\textbf{Personalization influences trust}}
When discussing improvements to AI tools that could increase trust, participants frequently mention personalization.
Health information varies across regions due to differences in medical systems between countries (P2, P20) and at an individual level based on personal symptoms or allergies (P14). To increase trust, our participants want ChatGPT to tailor its answers by considering these regional and individual differences. For example, act as a mediator to facilitate appointments with doctors, and 
proactively inquire about personal situations like a human doctor (P2, P14).
\begin{quote}
    \textit{``(ChatGPT) will ask you some relevant questions about your symptoms and maybe then, it can also find more specific information for you. Maybe it's a different story if you're a man and you're 80 years old, than if you're a woman and you're 18 years old like that, maybe it would give you different information.'' - P14}
\end{quote}

\section{Study 2: Comparison of dissemination interfaces for health information seeking}

\major{
Study 1 revealed that participants placed higher trust in the conversational search of ChatGPT over traditional search by Google, with participants citing the conversational and ``human-like'' nature of the interaction as a key factor.
However, as this finding was limited to a text-based interaction, it remains unclear whether this trust stems solely from the LLM's content generation or if enhancing the ``human-like'' experience through voice or embodiment would further amplify (or hinder) the trust. 
Drawing on the MAIN model's insights on information modality heuristics, Study 2 moves beyond the comparison with Google to focus exclusively on LLM-powered conversational search.
By holding the information source constant (using the same LLM backend), we isolate the impact of \textit{dissemination interfaces} (i.e., text-based, speech-based, and embodied) to understand how CUIs modalities modulate trust in generative health information.
}

\subsection{Study Methods}


\subsubsection{\textbf{Participants}}
A power analysis using GPower~\cite{gpower} determined that at least 15 participants were needed to detect a medium effect size with an alpha level of .05 and 80\% power. 
\begin{table}[htbp]
\centering
\footnotesize
\renewcommand{\arraystretch}{0.93}
\begin{tabularx}{\columnwidth}{p{5cm} >{\raggedright\arraybackslash}p{6cm} >{\raggedright\arraybackslash}X}
\toprule
\textbf{Demographic} & \textbf{Categories} & \textbf{Numbers of Participants (\%)} \\
\midrule
Gender & & (N=20)
\\ 
 & Female & 16 (80\%)
\\ 
 & Male & 4 (20\%)
\\
\hline
Age & & 
\\
 & 18-24 & 16 (80\%)
\\
 & 25-34 & 4 (20\%)
\\
\hline
Education & & 
\\
 & Bachelor & 11 (55\%)
\\
 & Master & 9 (45\%)
\\
\hline
Professional Domain & & 
\\ 
 & Health and Medical Science & 2 (10\%)
\\
 & Business, Economics, and Law & 8 (40\%)
\\
 & Arts, Culture and Entertainment & 5 (25\%)
\\
 & Science, Technology, Engineering, and Mathematics & 2 (10\%)
\\
 & Other & 3 (15\%)
\\
\hline
Frequency of online health \\information seeking & & 
\\
 & Often & 2 (10\%)
\\
 & Sometimes & 15 (75\%)
\\
 & Rarely & 3 (15\%)
\\
\bottomrule
\end{tabularx}
\vspace{-0.6mm}
\caption{Characteristic of participants in Study 2.}
\label{table:demographic_interface}
\vspace{-7.8mm}
\end{table}
We recruited 20 participants (N=20) through the institute's recruitment system. Participants had to be fluent in English, and participation was voluntary. 
Informed consent was obtained 
prior to the lab sessions. Participants received compensation in accordance with institute guidelines. 
The study received approval from ethics and data protection committee of the institute (ID: XXX).

The demographic information of the participants is listed in Table~\ref{table:demographic_interface}.
Participants were aged between 18 and 34 years, with 80\% falling in the 18-24 age bracket. Participants came from a variety of professional fields: 40\% from Business, Economics, and Law, 25\% from Arts, Culture, and Entertainment, 10\% from Health and Medical Sciences, 10\% from STEM, and 15\% from other fields. Regarding educational backgrounds, 55\% had undergraduate degrees, and 45\% held postgraduate qualifications. 
As for online health information-seeking experience, 10\% frequently used online sources, 75\% occasionally searched online, and 15\% rarely or never used online resources. 


\subsubsection{\textbf{Dissemination interfaces}}
\begin{figure}[!ht]
    \centering
    \begin{subfigure}[t]{0.33\textwidth}
        \centering
        \includegraphics[height=4cm]{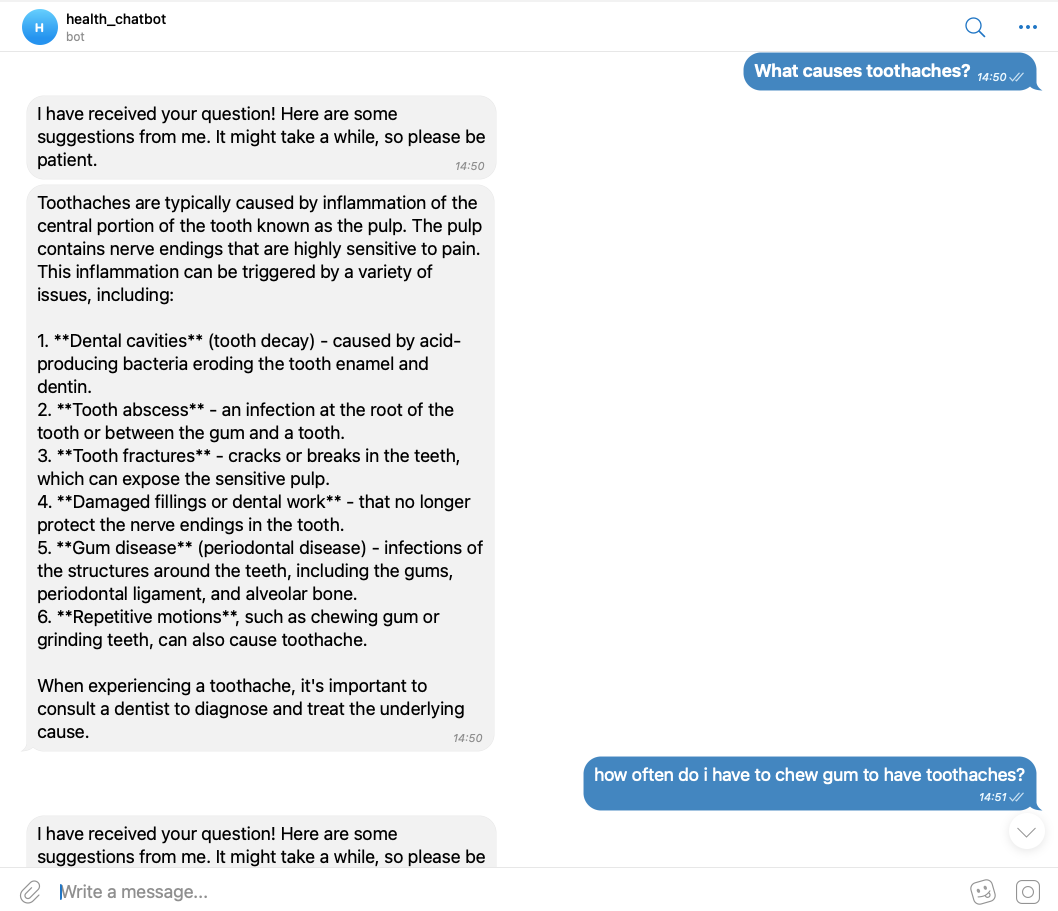} 
        \caption{text-based interface}
        \label{fig:image1}
    \end{subfigure}
    \hfill
    \begin{subfigure}[t]{0.33\textwidth}
        \centering
        \includegraphics[height=4cm]{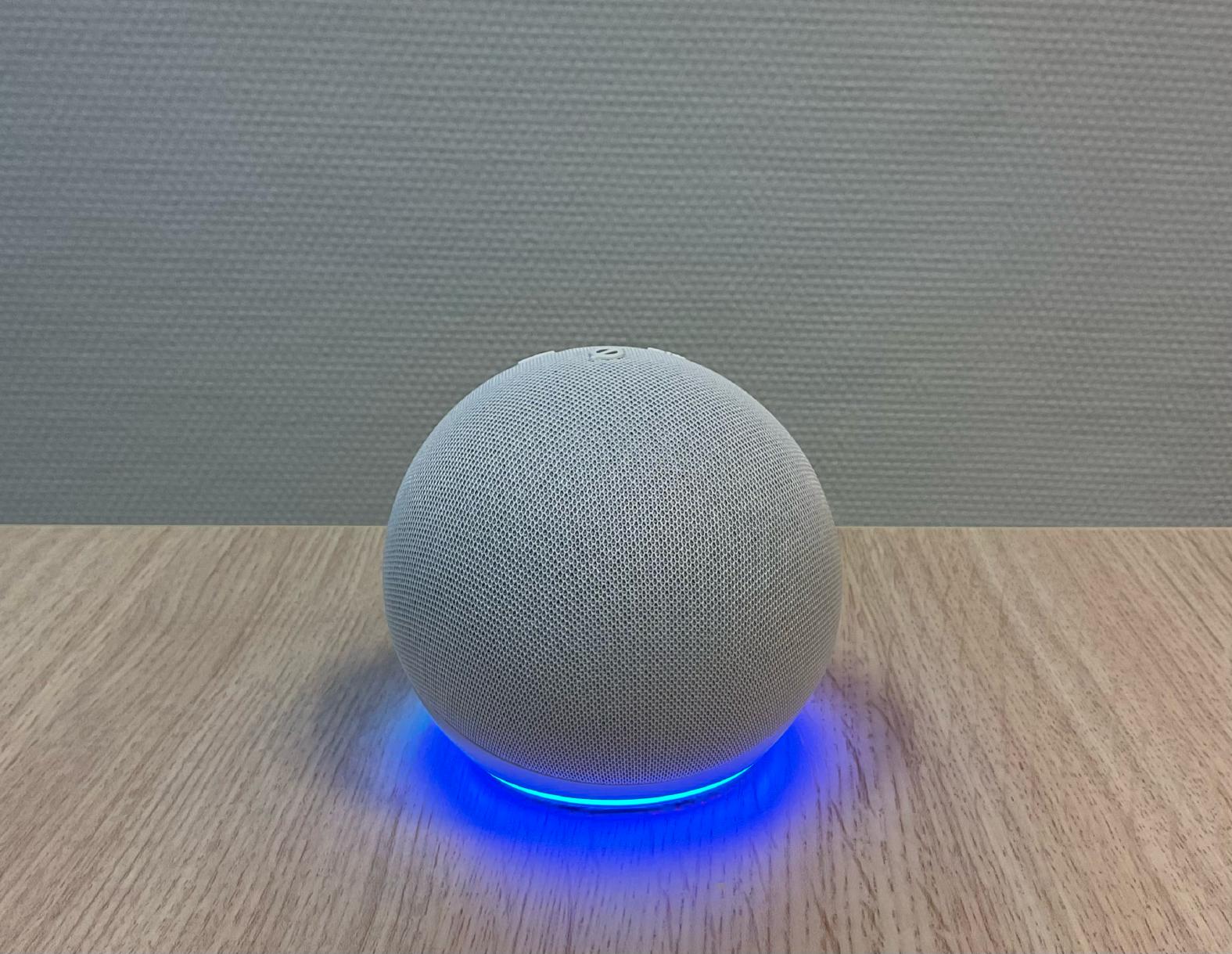}
        \caption{speech-based interface}
        \label{fig:image2}
    \end{subfigure}
    \hfill
    \begin{subfigure}[t]{0.33\textwidth}
        \centering
        \includegraphics[height=4.0cm]{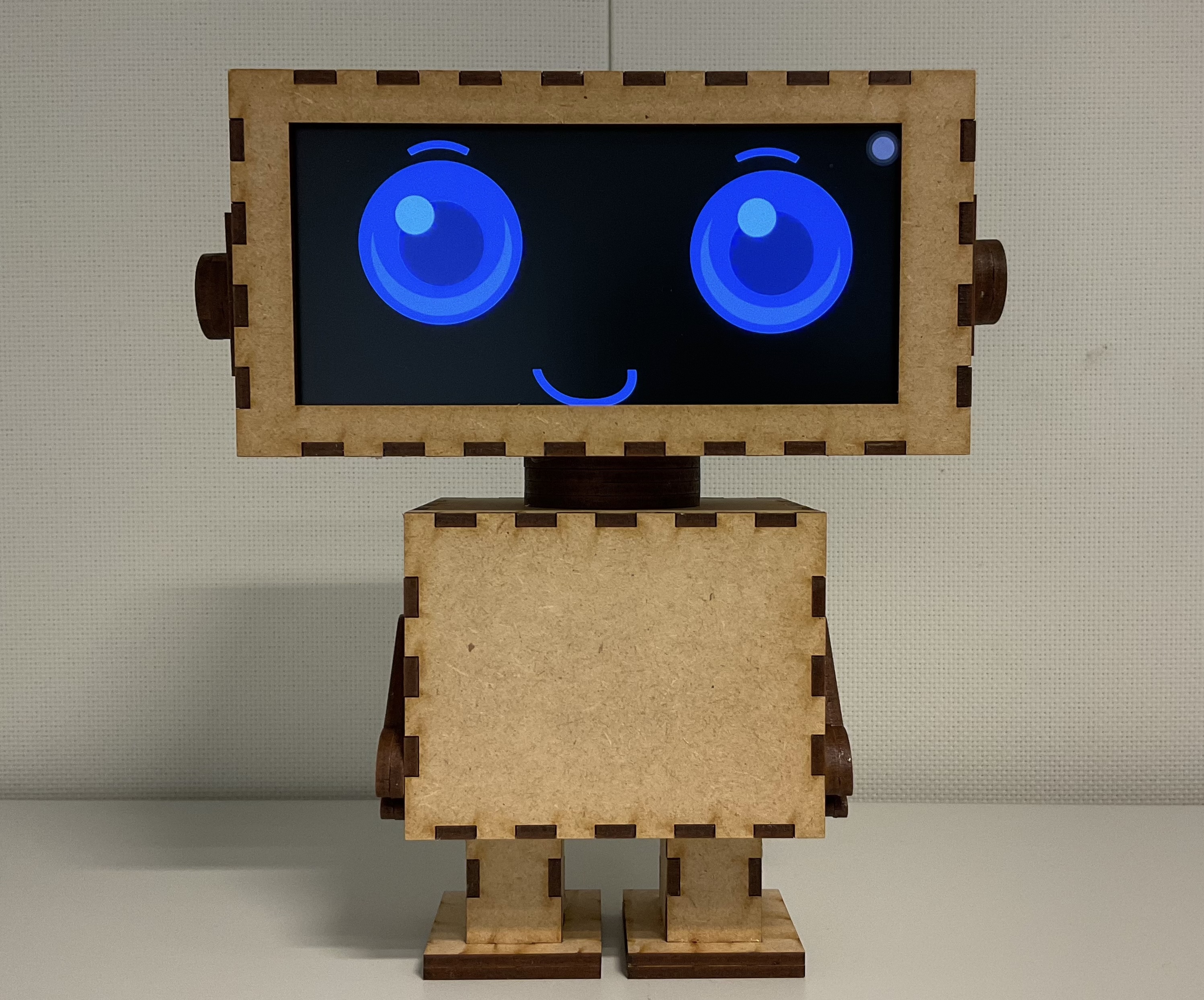} 
        \caption{embodied interface}
        \label{fig:image3}
    \end{subfigure}
    \vspace{-2.0mm}
    \caption{Three dissemination interfaces~\cite{trust_interface} were used in the lab study for participants to complete the health-related search tasks.}
    \label{fig:interfaces}
\vspace{-4.8mm}
\end{figure}

As shown in Fig~\ref{fig:interfaces}, Study 2 utilized three distinct LLM-powered conversational user interfaces: a text-based, a speech-based, and an embodied interface.
All three interfaces were powered by the same GPT-4o model~\cite{openai2023gpt4} to ensure generation consistency.
\vspace{-1.8mm}

\paragraph{Text-Based Interface:} 
We developed a web-based chat interface to facilitate straightforward, text-only interactions. This served as the baseline for assessing trust in information from LLM-powered conversational search.
\vspace{-2.8mm}

\paragraph{Speech-Based Interface:} 
Verbal interactions were enabled through an Amazon Echo Dot. 
To minimize brand association, wake words (e.g., ``Alexa'') were disabled, and visual feedback was limited to the device's standard light ring. 
This setup allowed for natural, spoken dialogue without physical embodiment.
\vspace{-2.8mm}

\paragraph{Embodied Interface:} 
\major{This setup featured a custom-built physical body with an internal speaker and a smartphone ``face'' displaying simple facial expressions. 
This embodied interface featured a physical, body-shape presence, enhancing the interaction with a tangible form.
The design intentionally utilized basic expressions to isolate the effect of physical presence, avoiding potential biases arising from hyper-realistic anthropomorphism or complex animations.}


\subsubsection{\textbf{Search tasks}}
The search tasks in this study were designed to align with those used in Study 1 to ensure consistency which is introduced in Section~\ref{search_tasks}. 
The tasks included 75 questions, categorized into three distinct types of health questions: general, symptom, and treatment-related questions.


\subsubsection{\textbf{Measures}}
The measures in this study were designed to align with those used in Study 1 to ensure consistency.
Besides, we asked the participants about their experience with the user interfaces selected and adopted two extra questionnaires to measure participants' eHealth~\cite{Health_Literacy} and AI literacy~\cite{ai_literacy} before the lab study.
During the lab study, we measured participants' perceived usability level using~\cite{UMUX}, and their intention to use the interface (adapted from~\cite{tam3}) for health information search after participants completed the interaction with each interface.


\subsubsection{\textbf{Procedure}}
The overview procedure of Study 2 is illustrated in Fig \ref{fig:procedure}.
Following informed consent, participants engaged in three search tasks per interface (text, speech, embodied) in a counterbalanced order, resulting in a total of nine tasks across the three interfaces. 
Participants were allowed to interact freely with each interface, including asking follow-up questions until they were satisfied with the answers.
We measured trust in information after each search task, and overall assessment of each specific interface, focusing on trust, usability, and their intention to use it in the future upon completing each interface condition.

Each lab session ended with a 15-minute semi-structured interview, where participants discussed trust factors, provided comparative evaluations, and offered suggestions for improvement of the interfaces.
The interview protocol is included as \textbf{Supporting Material}.


\subsubsection{\textbf{Data analysis}}

For the quantitative data, we first verified statistical assumptions using Shapiro-Wilk~\cite{SHAPIRO1965} (for normality) and Bartlett's~\cite{Arsham2011} (for homogeneity) tests. 
We then explored variable relationships via Pearson correlation~\cite{pearson}. 
To analyze trust differences across interfaces and types of search tasks, we employed a Mixed Linear Model (MixedLM)~\cite{mixedlm} controlling for usability based on the correlations analysis. 

For the qualitative data, we employed a thematic analysis~\cite{braun2012thematic} to examine interview responses, aiming to understand how trust in health information varies with different interfaces. To maintain consistency with Study 1, we followed the same analysis approach described in Section~\ref{qualitative_analysis}.

\subsection{Quantitative Findings: Lab Sessions}

\subsubsection{\textbf{Descriptive statistics}}

We conducted a descriptive analysis to examine factors in the study. 
Our findings showed that participants generally had a positive attitude towards technology, \major{with an average PPT score of 3.87 (SD=.33), indicating a favorable attitude towards using technology. eHealth literacy was moderately high, with a mean score of 3.68 (SD=.47), and AI literacy was also high, with a mean score of 3.70 (SD=.26).}


\begin{table}[!ht]
\centering
\renewcommand{\arraystretch}{0.94}
\begin{tabularx}{\columnwidth}{p{5.6cm} >{\raggedright\arraybackslash}X >{\raggedright\arraybackslash}X >{\raggedright\arraybackslash}X}
\toprule
\textbf{} & \textbf{Text-based \newline Mean (SD)} & \textbf{Speech-based \newline Mean (SD)} & \textbf{Embodied \newline Mean (SD)} \\
\hline
Prior experience (familiarity) & 3.60 (.75) & 2.35 (.88) & 1.60 (.75)
\\ 
Trust in information & 4.19 (.42) & 4.13 (.46) & 4.00 (.48)
\\ 
Trust in user interface &  3.73 (.37) &  3.57 (.46) & 3.56 (.44)
\\
Usability level of user interface & 4.05 (.43) & 3.55 (.67) & 3.55 (.63)
\\
Intention to use & 3.55 (.86) & 3.05 (1.07) & 2.70 (.95) 
\\
\bottomrule
\end{tabularx}
\vspace{-0.1mm}
\caption{Descriptive statistics of the measured variables in Study 2.}
\vspace{-7.0mm}
\label{table:descriptive}
\end{table}


\major{As shown in Table~\ref{table:descriptive},} usability level were highest for the text-based interface (M=4.05, SD=.43), compared to the speech-based (M=3.55, SD=.67) and the embodied interfaces (M=3.55, SD=.63), indicating that participants found the text-based interface easier to use, which correlated with their intention to use these interfaces for health information.
Regarding the familiarity level with the interfaces, participants were most familiar with the text-based interface (M=3.60, SD=.75), followed by the speech-based (M=2.35, SD=.88) and embodied interfaces (M=1.60, SD=.75). 
In terms of the intention to use the specific interface for health information search, participants most intended to use a text-based interface (M=3.55, SD=.86), followed by the speech-based (M=3.05, SD=1.07) and embodied interfaces (M=2.70, SD=.95). 

Table~\ref{table:descriptive} also summarizes mean trust levels in information and interfaces. 
Participants trusted the health information from the text-based interface the most (M = 4.19, SD=.42), followed by the speech-based (M=4.13, SD=.46) and embodied interfaces (M=4.00, SD=.48). Trust in the interfaces themselves showed a similar pattern: highest for the text-based interface (M=3.73, SD=.37), then the speech-based (M=3.57, SD=.46), and lastly the embodied interface (M=3.56, SD=.44).
Trust levels were relatively consistent across different types of search tasks in all three interfaces as shown in Figure~\ref{interface_fig_mean} (right), indicating similar trust in these types of search questions regardless of the interface used.

\major{Additionally, we assessed the recognition accuracy of the speech-based and embodied interfaces}, with each participant completing three tasks per interface, totaling 60 tasks each. For the speech-based interface, 10 tasks required one repetition and 3 required two, resulting in an accuracy of 83.3\%. The embodied interface had 7 tasks needing one repetition and 3 needing two, with an accuracy of 88.3\%. Follow-up queries were most common with the embodied interface (14 tasks), followed by the text-based (10 tasks) and speech-based interfaces (7 tasks).


\subsubsection{\textbf{Correlation analysis}}
\label{Correlation}

\begin{figure}[!ht]
\centering
\vspace{-2.0mm}
\includegraphics[width=0.99\textwidth]{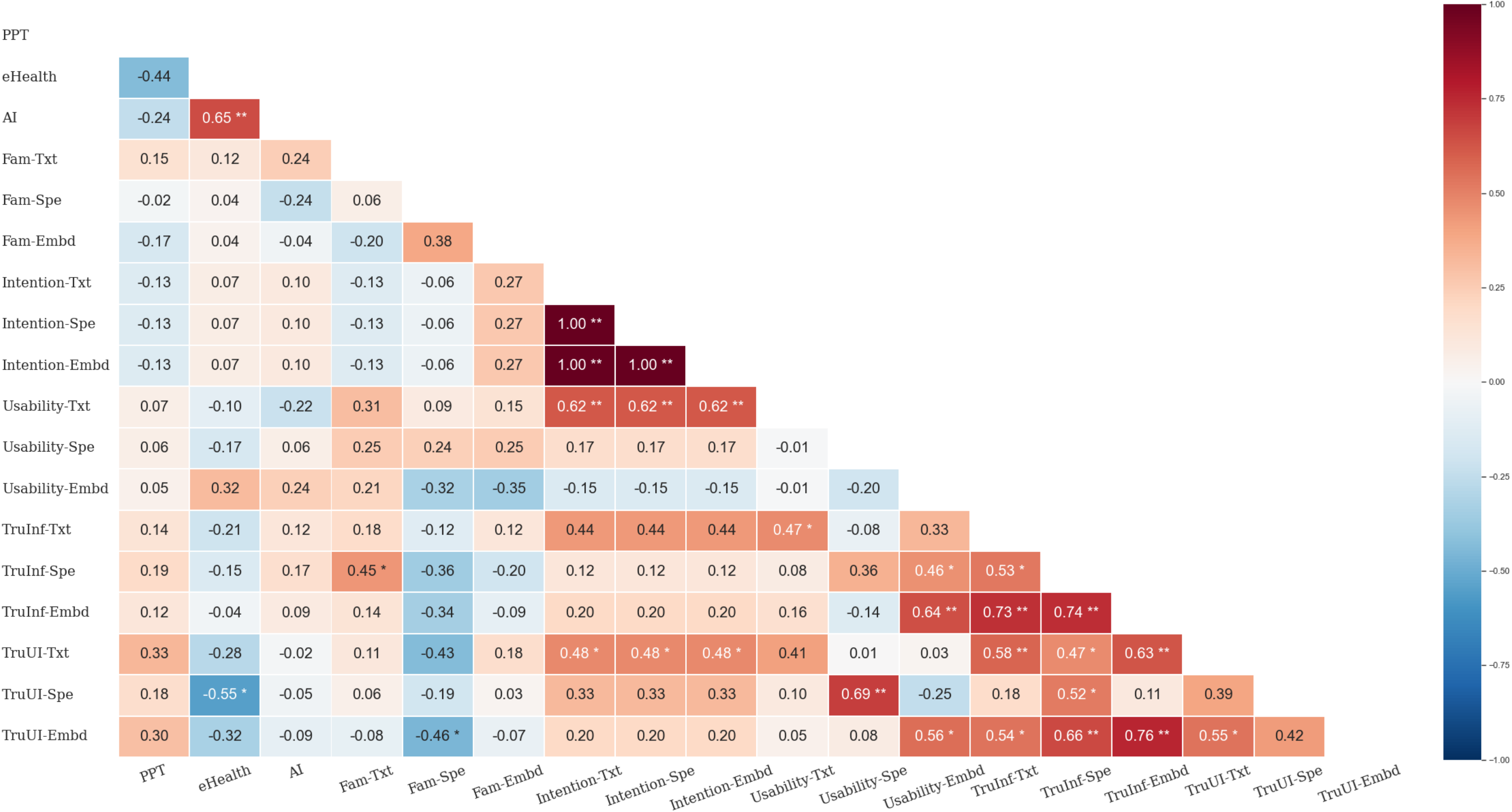}
\vspace{-0.2mm}
\caption{
Pearson correlation with key variables (**$p$<.01, *$p$<.05). Note: 
"PPT" means participants' propensity of trust in technology; 
"eHealth" and "AI" are participants' eHealth and AI literacy;
"TruInf" represents perceived trust level in health information obtained; 
"TruUI" means trust level in user interface; 
"Fam-" is the familiarity level of each UI; 
"Txt, Spe, Embd" represent three CUIs respectively. 
}
\vspace{-1.0mm}
\label{fig:interface_correlation}
\end{figure}

\major{Regarding RQ2b, which explores whether trust in the provided information correlates with trust in the specific dissemination interface,} the results of Pearson correlation analysis in Fig~\ref{fig:interface_correlation} revealed the relations among the key variables.

For the text-based interface, there was a significant correlation between trust in the information and usability level ($r$(20) = .47, $p$ < .05), as well as between trust in the information and trust in the interface itself ($r$(20) = .58, $p$ < .01).
In the case of speech-based interfaces, trust in the information was significantly correlated with trust in the interface ($r$(20) = .52, $p$ < .05). Trust in the interface, in turn, showed a significant correlation with usability level ($r$(20) = .69, $p$ < .01).
For embodied interfaces, trust in the information was significantly correlated with trust in the interface itself ($r$(20)=.76, $p$<.01) and usability level ($r$(20)=.64, $p$<.01). Additionally, trust in the interface was significantly linked to usability level ($r$(20)=.56, $p$<.05).
These results highlight a consistent pattern: across all types of interfaces, trust in health information was closely related to trust in the interface. Additionally, trust in the health information was strongly connected to the usability level, particularly for text-based and embodied interfaces.


\subsubsection{\textbf{Human trust in information differs by dissemination interfaces}}

The correlation analysis identified significant relationships between usability, trust in the interface, and trust in the information provided. Following these findings \major{and to address RQ1b, which examines how different dissemination interfaces (text, speech, embodied) influence perceived trust,} we applied a Mixed Linear Model (MixedLM)~\cite{mixedlm} regression to control for variables such as usability, trust in the interface, and familiarity in each interface.

\begin{figure}[!ht]
    \centering
    \begin{subfigure}[t]{0.99\textwidth}
        \centering
        \includegraphics[height=4.06cm]{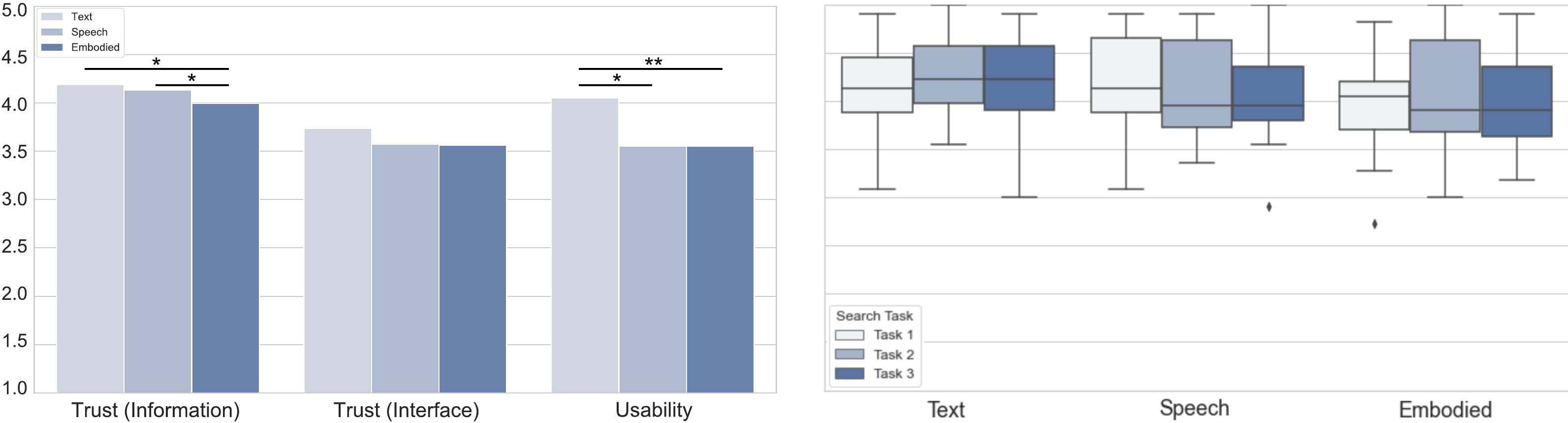} %
        \label{interface-mean-difference}
    \end{subfigure}
    \vspace{-1.6mm}
    \caption{
    Left: Mean of perceived trust scores in health information and in the interface, mean of usability level, and level of intention to use each interface for health information searches. 
    Right: Median of trust scores in health information for three types of search tasks across three user interfaces.
    (**$p$<.01, *$p$<.05).
    }
    \vspace{-1.6mm}
\label{interface_fig_mean}
\end{figure}

\major{As shown in Figure~\ref{interface_fig_mean},} the MixedLM indicated significant differences of trust in health information disseminated between text-based and embodied interfaces ($\beta = .189, p = .035$) and between speech-based and embodied interfaces ($\beta = .180, p = .002$). These differences were largely influenced by usability which can significantly predict the trust in the information ($\beta = .187, p = .004$). 
There was no significant difference in trust levels across various types of search tasks (\( p = .510 \)), suggesting that trust remains consistent regardless of the type of health-related question.
A mediation analysis further showed that usability fully mediates the relationship between interface type and trust in information, with a significant indirect effect ($\beta = .154, p < .001$) and a non-significant direct effect ($\beta = .098 ., p = .277$).
These findings highlight that usability and trust in the interface are crucial factors affecting trust in information.

\subsection{Qualitative Findings: Semi-Structured Interviews}

\major{This section outlines themes from the interviews to reveal the underlying factors shaping trust across different modalities (RQ3b).} 
First, we explore \textit{``what factors impact trust perceptions in health information disseminated by different interfaces''}. 
Second, we explore \textit{``how to improve LLM-powered CUIs for trustworthy health information seeking''}.


\subsubsection{\textbf{What factor impacts trust perceptions in health information disseminated by different interfaces?}}


\paragraph{\textbf{Prior experience and familiarity with the interface influence trust in information.}}
Prior experience and familiarity with the interface are key factors in establishing trust in the information, as supported by our quantitative findings. Interviews indicated that participants' trust was primarily influenced by their familiarity with different interfaces and their habits in seeking online health information (P2-3, P5, P8-10, P11-12, P14, P16). For example, participants who frequently used Google or professional health websites found text-based interfaces more familiar and trustworthy. Moreover, the resemblance of text-based interfaces to widely-used social media platforms like WhatsApp and their similarity to consulting human health professionals or reading professional literature (P3, P6, P7) further increased trust in these interfaces.
\begin{quote}
    \textit{``I think just about the presentations because the way I chat with the chatbot is more reliable because it provides me the information like I used to, uh, familiar with.'' - P2}
\end{quote}


\paragraph{\textbf{Usability of interfaces greatly impacts trust in information.}}
As indicated by our quantitative findings, particularly with text-based interfaces where usability mediated trust. Participants favored text-based UIs for their simplicity and ease of use (P1, P4-P7, P9-11, P18). While trust in text-based applications has already been established with the rise of LLMs, usability is crucial for building trust in less familiar speech and embodied interfaces (P1, P13, P18, P20).
\begin{quote}
    \textit{``I think this is more likely to use it. [...], because more likely to use anything like adding supplementary questions, you can sort of build trust on the information.'' - P18}
\end{quote}


\paragraph{\textbf{Information presentation style impacts the trust in information delivered by interfaces.}}
The way information is presented significantly influences trust, especially in speech and embodied interfaces. Although the information source was consistent across all interfaces to prevent content bias, participants noted that while numbered formats work well for text, they may cause information overload in vocal formats, leading to cognitive overload and forgetfulness (P5-6, P10, P12, P20). Adjusting presentation styles, such as using storytelling or summarizing key points at the start of interactions can improve engagement and trust in these interfaces (P5, P10, P13).
\begin{quote}
    \textit{``It is better to present quite an amount of information more like storytelling I would say, [...] Just like an essay like instead if you don't put it in bullet points but you say like you use connection words.'' - P10}
\end{quote}


\paragraph{\textbf{Information modality influences how information is processed, which in turn affects trust.}}
The way people process different information modalities significantly affects their trust. Participants found it easier to process text than vocal information, as reading required less effort than listening (P5, P10, P12, P20). This ease of processing allows users to focus more on the content, which enhances trust, \textit{I prefer to read book paper-wise rather than listening to it (P12).}
Text-based information also mitigates issues related to context retrieval and timeliness found in speech, enabling users to cross-compare information within the same interface, thus deepening their understanding and building trust incrementally (P1, P4, P5, P7, P9, P11, P14, P16). Additionally, text-based interfaces allow easy sharing of information with health professionals or friends (P11) and facilitate detailed exploration and clarification of complex medical terms.
\begin{quote}
    \textit{``If I don't even know what I missed, don't know what kind of questions to follow up, whereas text one, I can see everything and know exactly what I want to ask. Because I can see and go back and refer to the information.'' - P7}
\end{quote}



\noindent Additionally, participants expressed that the embodied interface requires processing both verbal information and non-verbal cues like facial expressions (P3-P5, P7-P8, P11-12). This extra information processing can raise cognitive load~\cite{Robb2023SeeingET}, further affecting users' trust perception of the delivered health information.
\begin{quote}
    \textit{``If I'm asking a question (to embodied and speech interfaces), I have to listen because I don't know what's coming next. [...], you had to actively listen. And if you missed one part, then you cannot go back.'' - P5}
\end{quote}


\paragraph{\textbf{Human-like features introduce additional considerations for trust.}}
\label{sec:human-like2}
Participants reported that compared to text-based and speech-based interfaces, a physical embodiment gave a stronger sense of human interaction, leading to greater engagement and more follow-up questions (P3-4, P10). Our quantitative finding supports that participants asked more follow-up questions with the embodied interface than the other two.
However, physical embodiment introduces factors that influence trust perception, such as the voice and appearance of the interface (P1, P3-P8, P11). 
Besides, the embodied interface can offer advantages similar to interacting with health professionals, such as observing symptoms and providing personalized advice, which text or speech interfaces lack. However, trust in human-like communication can vary. Some participants may distrust humans, leading to negative perceptions of human-like interactions.
These benefits also present challenges to privacy concerns associated with its use and personal data leakage (P3, P5, P7, P11, P13-14). As P3 expressed worry about data sharing:
\begin{quote}
    \textit{``Maybe it (embodied) can recognize my sound also track some information and like store the information in which may cause some like security problems maybe.'' - P3}
\end{quote}
Overall, participants preferred embodied interfaces that balanced human-like and robotic features, favoring those that appeared neutral yet professional, with well-matched voices and movements. 
\begin{quote}
    \textit{``It is like the uncanny valley, I don't know how to say that, but there is the balance to bring, not be in the horror things but also (the interface) can be advanced.'' - P3}
\end{quote}


\subsubsection{\textbf{How to improve LLM-powered CUIs for trustworthy health information seeking?}}


\paragraph{\textbf{Enhancing features to improve information credibility.}}
Participants were primarily concerned about the credibility of source information (P1, P4-5, P7, P9-10, P13-16), with distrust often stemming from a lack of references or the use of fake ones (P7, P9). They suggested adding features that link to original sources, allowing users to verify information, thereby increasing trust. 
\begin{quote}
    \textit{``I think if they can tell me this was taken from this XYZ source and cross reference with XYZX number of sources, I would probably trust it.'' - P7}
\end{quote}
Additionally, CUIs can provide feedback that shows understanding of user queries, which can enhance trust in both the interface and the information provided (P5, P8). The CUIs can proactively ask contextually relevant follow-up questions (P2, P10) when they are not fully confident in understanding user's queries.
\begin{quote}
    \textit{``It could be nice that you have like a light that's seeing when it's ready to get a response already. To get a question, and when it's talking, you have a different color or something so that you're assured that it gets your head has kept your question right or has captured your answer.'' - P8}
\end{quote}
For embodied interfaces, participants recommended enhancing the interactive experience by synchronizing vocal interactions with more expressive features, as this could strengthen better relationship between the user and interface and further build trust (P1, P3-8, P11, P16).


\paragraph{\textbf{Enriching interaction by integrating multiple information modalities and improving trust.}}
Participants recommended incorporating multiple information modalities within a single interface. They felt that adding images or videos to text could improve comprehension, especially in health-related contexts (P4, P9). For instance, using images to show specific pain areas would enhance a text-based interface. In embodied interfaces, combining text, images, and videos was viewed as beneficial for boosting both trust and usability. This multi-modal approach could accommodate diverse searching styles and preferences, thereby enhancing the effectiveness of the health information delivered.
\begin{quote}
    \textit{''It would be best I can do a bit mixed with text and speech like type my questions and talk to the interface at same time. [...] If there is a screen in front of the embodied robot to display the word is gonna give more trust feeling.'' - P4}
\end{quote}


\paragraph{\textbf{Personalize information toward individual needs.}}
\major{Echoing the qualitative findings from Study 1 that participants prefer that LLM's responses are tailored to users}, participants in Study 2 also emphasized that health information should account for individual situations, such as personal symptoms or allergies, as well as regional differences, like variations in medical systems across countries (P2). They identified the generic nature of the information provided by interfaces as a barrier to trust and expressed a need for more personalized responses, particularly from embodied interfaces (P1-3, P7, P10-11).
The trust would be significantly improved if interfaces could proactively inquire about personal symptoms to provide tailored advice (P2) and ask contextually relevant follow-up questions (P2, P10).
\begin{quote}
    \textit{``I think my feeling of trust will increase because the information given to me is more personalized and it might be something that I can't even identify myself and therefore don't know to add into the speech or the text to search if the bot can see things that I cannot.'' - P7}
\end{quote}



\section{Discussion}

\major{The results from both studies reveal that perceived trust in health information varies significantly across search agents and dissemination interfaces, highlighting a shift from the traditional ``retrieval'' model of search engines to the emerging ``generator'' model of LLMs with the advanced multi-modal CUIs. 
Following this, we discuss the key findings and their design recommendations based on the observations from both of our studies.}


\subsection{Divergent Trust Patterns: From Search Agents to Dissemination Interfaces}

In Study 1, participants showed a higher level of trust in ChatGPT compared to Google. 
Our findings showed that ChatGPT's conversational style and rapid, personalized responses made users feel heard and understood~\cite{Rheu}, which can be perceived as a sign of expertise and relevance~\cite{Zhongxuan, hai_1}. 
Its intuitive and user-friendly interface may further contribute to perceived credibility~\cite{21, 8}. 

\major{Beyond overall trust levels, the relationship between agent trust and information trust varied by types of search agents.
For Google, we observed 
a significant correlation between trust in health information provided by Google and trust in Google as a search engine. This likely reflects longstanding familiarity with traditional search engines, where users have developed a good relationship with Google, leading to confidence in both the agent and the information it provides~\cite{Haque, hai_7}. 
In contrast, no significant correlation was found between trust in ChatGPT and trust in the health information it generated, suggesting that people may separate their trust in ChatGPT from their trust in the content it produces. Despite ChatGPT's advanced capabilities, users may still be in the process of developing trust relationships with conversational search, particularly in newer configurations such as ChatGPT search~\cite{gpt_search} 
}

\major{
This divergent trust pattern in health information (as defined in Section~\ref{search_tasks}) can be explained by differences in how information sources are structured and perceived.
In the Google condition, the structural separation between the search agent Google and the external medical websites allows users to clearly distinguish the information provider and evaluate source credibility independently.
In contrast, conversational search collapses this distinction. We refer to this pattern as \textit{source dissociation}, in which trust in the conversational agent does not translate into trust in the health information it generates.
Thus, no significant correlation was observed between inherent trust in the conversational agent and trust in the information output.}
\major{
Qualitative findings help explain this pattern. Our participants evaluated synthesized responses primarily based on intrinsic presentation rather than the agent's reputation. As discussed in Section~\ref{factors_agents}, credibility judgments were shaped by prior experience (P11), logical flow, and professional language (P1, P2, P17).
In the absence of an explicit source, users relied on trust heuristics defined in the MATCH model~\cite{responsible_ai} as a proxy for trustworthiness and credibility, effectively decoupling trust in the information from the trust in the agent.
}

Regarding dissemination interfaces, \major{our data from Study 2} challenges the assumption that higher modality richness (e.g., voice or avatars) linearly enhances trust. 
\major{Despite the MAIN model's suggestion that modality and social cues may increase credibility~\cite{Sundar2007TheMM}, our quantitative trust ratings showed a significantly higher trust in text-based interface over speech-based or embodied ones.
While speech-based and embodied interfaces provided more natural interactions, our participants favored text-based interfaces for health inquiries because they allowed for easier and more precise cross-referencing, a critical factor when seeking accurate medical advice.
Besides, the familiarity of text-based interaction also appears to outweigh the value of richer modalities, particularly in high-stakes health scenarios.
Our quantitative analysis further supported these findings, showing that trust in an interface directly influenced trust in the information it provided.}
This aligns with established research~\cite{ti, 9, Zhang1, 14} confirming that source credibility is a key factor in how users evaluate health information. 
This highlights the importance of designing LLM-powered tools that not only deliver accurate health information but also foster user trust with such tools.

\major{Lastly, across both studies, we found no significant main effect of task type on trust ratings.} 
This indicates that the trust patterns described above, preferring conversational agents but text-based CUIs, are robust and consistent across different categories of health inquiries, suggesting a generalized behavior rather than a task-specific anomaly.


\subsection{\major{Factors of Trust in LLM-powered Conversational Search for Health Information Seeking}}

\major{We identified the key factors that drive the trust patterns observed above for LLM-powered conversational search by synthesizing our qualitative insights from both studies.}

\subsubsection{\textbf{The Role of Search Autonomy and Familiarity}}

Our qualitative data identifies two factors that influence trust in health information: the user's \textit{search autonomy} (moderated by prior knowledge) and reliance on \textit{familiarity}.

\textbf{Search Autonomy and Prior Knowledge.}
Search autonomy, the degree of control users exercise over the information seeking process, emerged as a key factor shaped by the user's prior knowledge of health queries.
\major{Our interview data reveal that participants preferred Google because it supported an autonomous ``searching'' process. They valued Google’s ability to cross-reference multiple sources to verify consistency, thereby enhancing their confidence in the information~\cite{t13}, and they viewed this transparency as a prerequisite for trust.
However, participants with limited prior knowledge favored the low-autonomy, high-synthesis approach of ChatGPT. 
As noted by our participants from Study 1, LLM served as a ``knowledge scaffold'', transforming complex medical jargon into easier-to-understand summaries. 
Our participants viewed ChatGPT as a convenient starting point, offering initial insights that guide further detailed searches.
This aligns with Xu et al.~\cite{xu2023chatgpt},} suggesting that for novices, the efficiency of a direct answer outweighs the benefits of deeper verification.
Thus, prior knowledge serves as an internal trust calibrator, reducing the need for extensive autonomous search. In the LLM context, search autonomy is redefined as the ability to ask iterative follow-up questions, a crucial feature noted by our participants who prioritize verification. 
This highlights a divergence in behavior: higher prior knowledge mitigates the reliance on search autonomy, whereas the drive for information validation reinforces it.

\textbf{Familiarity.}
Beyond search autonomy, trust was heavily influenced by familiarity level and their prior experience, as supported by previous studies~\cite{ti, Harrington, Luria, I_Can_Do_Better, raju1980product}. 
\major{In Study 1, the strong correlation between trust in Google (the agent) and its information can be attributed to a ``legacy effect''. 
Participants described their trust in Google as a product of higher familiarity, echoing prior work~\cite{kuang2023collaboration}, whereas trust in ChatGPT is still nascent. 
In Study 2, this familiarity bias strongly favored text-based interfaces.
Our qualitative data indicated that participants trusted text interfaces not necessarily because they were technically superior, but because they aligned with their habitual methods of health information seeking.}
Speech and embodied interfaces, despite their novelty, introduced extra cognitive load that disrupted this familiarity.
These findings underscore that fostering trust in emerging conversational agents and interfaces requires more than technological advancement. It demands a design strategy that builds familiarity by aligning new technologies with users' established search habits for online health information.



\subsubsection{\textbf{Usability of Conversational Search}}


\major{Grounded in the established trust model~\cite{ti} and prior work~\cite{ellis1993comparison}, our findings confirm that usability (e.g., ease of use) is not merely a functional attribute but a fundamental prerequisite for trust formation. 
However, our qualitative insights reveal that the definition of ``usability'' shifts depending on the complexity of the conversational search.}

\major{\textbf{Efficiency vs. Transparency (Study 1).}}
For search agents, usability influenced trust through two distinct mechanisms: \textit{cognitive ease} and \textit{process transparency}.
Participants who favored ChatGPT equated usability with efficiency, the ability to obtain quick answers and initial insights. 
This aligns with Xu et al.~\cite{xu2023chatgpt}, showing that users spend less time on ChatGPT for similar tasks compared to Google, which indicates a preference for its efficiency. 
Conversely, users who preferred Google defined usability as transparency, the ability to easily cross-reference sources. Our qualitative data indicates that for these users, the ``streamlined'' nature of ChatGPT was perceived not as usable, but as opaque, whereas Google’s interface facilitated a transparent verification process essential for high-stakes health queries.

\major{\textbf{Interactive Complexity of Multimodal CUIs (Study 2).}}
Regarding dissemination interfaces, prior research suggests that multi-modal interfaces such as embodied and speech interfaces increase engagement by integrating visual and auditory elements next to text alone~\cite{multimodal_ir_1, multimodal_ir_2, multimodal_ir_3, multimodal_ir_4, Robb2023SeeingET}. 
\major{Besides, Deldjoo et al.~\cite{multimodal_ir_3} indicate that multi-modal systems can facilitate more comprehensive and immersive comprehension of content by stimulating multiple senses. As evident by the MAIN model from~\cite{Sundar2007TheMM}, modality can influence the perception and trust of the information.
However, our participants from Study 2 interviews reported that this added richness but introduced unnecessary cognitive load~\cite{ethics-4, Robb2023SeeingET}.
Processing visual and auditory cues required more mental effort than simply scanning text. 
Our participants expressed that the increased complexity of multi-modality can impact usability and, subsequently, trust perception.}
Our participants also mentioned that multi-modal interactions may feel more engaging and human-like, they can introduce opportunities for errors, echoing the prior studies, such as misinterpreting tone in speech~\cite{voice-1, voice-2, voice-3, voice-4} or synchronization issues in embodied interactions~\cite{Embodiment-1, Embodiment-2, Embodiment-3, Embodiment-4, Embodiment-5}.
As noted by participants, the straightforward nature of text-based interfaces reduced such opportunities, fostering greater transparency and reliability in information.

Furthermore, our qualitative data highlight a lower tolerance for technical inconsistencies in multi-modal interactions.
Participants noted that minor errors, such as a synchronization lag in the embodied avatar's expressions, can immediately break the illusion of competence, leading to skepticism~\cite{Embodiment-1,ethics-4}.
In contrast, text-based interfaces offered a robust medium. 
Therefore, the usability design should be prioritized to foster trust in digital tools for health search.



\subsubsection{\textbf{Perceived Anthropomorphism and Personalization in LLM-powered Health Information Seeking}}

\major{Our qualitative analysis (Sec~\ref{sec:human-like1} and \ref{sec:human-like2} ) reveals that trust in conversational search is heavily influenced by two complex factors introduced by the advanced LLMs: the perception of anthropomorphic features and the depth of personalization. 
}

\textbf{Human-like Features vs. Human Identity.} 
\major{Prior research indicates that the anthropomorphism of LLMs influences user trust~\cite{PELAU2021106855, Pelau, ChatGPT_therapist, t11, Korteling}. Specifically, their conversational nature fosters intuitive information delivery, while role-playing capabilities~\cite{roleplay-2,roleplay-1}, such as simulating a doctor's persona to enhance perceived relevance and credibility~\cite{hai_6}.
Cai et al.~\cite{Personal_Characteristics} find that mirroring the style of human advisors increases user attention and comfort. 
}
\major{Interpreting our findings through the lens of the MAIN model~\cite{Sundar2007TheMM} regarding information modality and anthropomorphism, we observe that while our participants appreciate the functional benefits of such human-like features, they remained cautious toward deceptive anthropomorphism.}
\major{Participants indicated that trust is enhanced when LLMs adopt a human-like conversational style, specifically, to break down complex medical jargon into manageable pieces rather than overwhelming users with lengthy responses. 
Moreover, incorporating empathy, encouragement, and understanding in LLM interactions can further foster deeper user engagement and trust~\cite{emo-emp-1,emo-emp-2,emo-emp-3}, a point that our participants echoed.}

\major{
However, Troshani et al.\cite{Anthropomorphism-6} and Placani\cite{Anthropomorphism-8} argue that making LLM-powered AI highly human-like in appearance does not necessarily enhance trust, particularly in health contexts where users often prefer neutral, clear information~\cite{gg_gpt_2}.
Placani~\cite{Anthropomorphism-8} and Festerling et al.~\cite{Anthropomorphism-9} suggest that using familiar language, a reassuring tone, and a patient-centered approach, similar to traditional healthcare, can be more effective for fostering trust~\cite{I_Can_Do_Better, Luria}.  
Our findings support that ``human-like'' does not imply a desire for AI to pretend to be human. Human-like features can evoke discomfort or skepticism due to ``uncanny valley'' effect~\cite{wikipediaUncannyValley}. 
Our findings reinforce this distinction: participants appreciate human-like \textit{communication} but are concerned about the pretense of human \textit{identity}. 
In Study 2, participants emphasized the necessity of source transparency. 
Clearly identifying the source as an AI allowed users to set realistic expectations and critically assess the content, aligning with prior work on source identification for trust~\cite{Anthropomorphism-8, Zhang1, chin2002patients,trust_physio}. 
Thus, effective trust-building requires a balance: leveraging the empathy and clarity of human-style conversation~\cite{emo-emp-1, emo-emp-3} while maintaining explicit transparency about the agent's artificial nature to avoid confusion or skepticism~\cite{I_Can_Do_Better, gg_gpt_2}.
}


\textbf{The Trade-off Between Personalization and Privacy.} 
Our findings also identify personalization as a double-edged sword: it boosts engagement and utility-based trust, yet simultaneously triggers privacy-related distrust.

\major{
On one hand, our participants expressed that trust is built when LLMs demonstrate ``contextual empathy'', adapting their tone to the user’s specific symptoms or emotional state. 
Unlike static search engines, the LLM’s ability to handle follow-up questions and interpret complex, multi-faceted health queries with contextual depth, allowing for responses that are tailored to individual needs, which makes users feel ``being heard''. 
This corroborates Sillence et al.~\cite{SILLENCE20071853}, who argue that people are more likely to trust personalized responses when seeking health information online.
This is particularly valuable in health contexts, where advice tailored to specific symptoms or emotions can make tools appear more empathetic and supportive, 
maintaining continuity in interactions and improving user experience~\cite{t12}.
}

\major{
On the other hand, as our participants revealed in the interviews of Study 2, the depth of contextual interaction can also raise significant anxiety regarding the privacy of data handling~\cite{privacy-1, privacy-2}. 
As the interaction becomes more personalized (and thus requires more sensitive personal data input), participants expressed concerns about data leaks and surveillance. 
This was particularly evident regarding multi-modal features of CUIs, as our participants feared that voice or embodied interfaces might engage in ``constant listening'' or visual monitoring~\cite{Capture-1, Capture-2} without informed consent~\cite{proactive_ca}, increasing the potential for misuse or breaches of personal data.
This tension highlights a critical insight: trust in personalized health AI is conditional on data safety. As noted by Bansal et al.~\cite{Bansal}, trust depends not only on the content credibility but also on how data is managed. 
While users desire the benefits of tailored health advice, our interview data suggests that robust data protection and transparent privacy protocols~\cite{Ethical, ethics-2, t13, responsible_ai} are a non-negotiable prerequisite for maintaining trust in these personalized interactions in highly sensitive health contexts.
}


\subsection{
\major{Design Recommendations for LLM-powered Health Information Systems}
}

\major{Grounded in our qualitative evidence and prior literature, we propose three design recommendations for future development of trustworthy LLM-powered health tools and systems.}

\textbf{Bridging the gap between search and verification.} 
\major{Our findings from Study 1 indicate that users often use LLMs as a starting point rather than a final authority, actively seeking to verify information with multiple information sources. 
Current LLM-powered CUIs, however, often present generated answers as absolute truth, forcing users to leave the platform to verify facts. 
To support this user-driven verification behavior related to the search autonomy, designs should act as a ``bridge'' rather than just an answer generator. 
This could involve integrating clickable provenance where users can trace specific medical claims back to source documents directly within the chat flow. 
Furthermore, when an LLM detects high-stakes medical queries, it could proactively prompt users with verification cues (e.g., ``This is a complex condition; would you like to see clinical guidelines?''). 
By supporting rather than replacing the user's autonomy in information seeking, LLMs can shift from being viewed as potentially hallucinatory chatbots to reliable search assistants, especially in sensitive health contexts.}

\textbf{Designing for functional anthropomorphism with explicit identity.}
\major{Our study reveals a critical distinction: participants trust the \textit{patterns} of a human advisor (empathy, reflective) but distrust the \textit{visual pretense} (avatars, ambiguous identities). 
This suggests a design shift from full anthropomorphism to functional anthropomorphism. 
Designers should prioritize replicating the \textit{conversational style} of a health consultation, such as breaking down complex diagnoses, offering reassurance, without attempting to visually or structurally impersonate a human doctor. Interfaces should avoid the ``uncanny valley''~\cite{wikipediaUncannyValley} by maintaining an explicit identification for AI (e.g., using clear labeling~\cite{trust_physio}) while retaining the warm, explanatory tone that users value.}
It aligns with the need for source transparency~\cite{Zhang1} while delivering the cognitive comfort of human-like advice~\cite{Personal_Characteristics}.

\textbf{User-aware calibration.}
\major{Trust in health AI relies on the system's ability and modality to align with user's specific needs and mental states. 
Our participants highlighted that a ``reassuring tone'' increased their trust. This points to a unique opportunity for LLMs to function not just as informational tools, but as agents of emotional regulation. Design recommendations could be implementing affect-aware calibration~\cite{trust_physio}, where the system detects signs of user distress or high-stakes symptoms and prioritizes ``affective safety'' in its output.
By balancing clinical accuracy with emotional intelligence, LLM-powered CUIs can mitigate the ``cyberchondria''~\cite{cyberchondria} often associated with online health seeking, creating a safer environment for health inquiry.
In addition, to address varying levels of prior knowledge (e.g., health literacy), designs should tailor to different users employing dynamic scaffolding~\cite{scaffolding}. 
By implicitly detecting a user's expertise (e.g., via their use of medical terminology), systems can adapt their response complexity. 
Moreover, the system or the CUIs can adapt their output into various modalities based on user states and literacy level, as suggested by our participants (P4, P9), to incorporate multiple information modalities to improve trust.}

\textbf{Privacy-preserving personalization.} 
\major{The tension between the utility of personalized advice and fear of data surveillance emerged as a key barrier to trust of LLMs in health contexts. 
While participants valued contextual and tailored responses, they simultaneously expressed concerns about constant monitoring and listening. 
In response, LLM-powered interfaces could introduce features such as incognito health modes or explicit forgetting mechanisms that allow users to disable memory retention for sensitive queries selectively. In addition, different modalities of privacy indicators should be explored~\cite{schaub2015design,Bansal}. For example, visual cues that signal when prior symptom data are being used to generate a response could help make personalization processes legible, transforming opaque data practices into transparent and accountable design features. 
Future systems should move beyond binary consent models (accept/reject) toward more granular and context-sensitive privacy controls.
In line with theories of contextual integrity, health-related interactions should be designed with attention to appropriate information flows, including the roles of information providers and recipients, transmission principles, and types of data involved~\cite{nissenbaum2004privacy}.
}

\section{
\major{Limitations and future work}
}

While our study provides valuable insights into the differences in trust in health information from distinct search agents and across CUIs, it is important to acknowledge its limitations for a more nuanced understanding of trust.

First, our research focused solely on Google and ChatGPT as representatives of traditional search engines and conversational search, respectively. While these are currently the most prominent examples, we recognize the emergence of hybrid approaches, such as Bing~\cite{bing} and ChatGPT Search~\cite{gpt_search}, which combine web browsing with LLM capabilities. 
Nevertheless, our study serves as a foundational step in unraveling insights into human trust and reliance on information provided by traditional and LLM-powered search agents. 
In addition, our study focuses on health-related information. The differences in trust perceptions may vary when considering other types of information or contexts, and this specificity could limit the generalizability of our findings. 
Future work should investigate how emergent hybrid technologies that blend browsing with generation influence trust, while also validating these perceptions across diverse information domains to ensure the broader generalizability of our findings.


Second, our research design utilized a lab study followed by semi-structured interviews. While this method balanced the need for in-depth insights with privacy concerns regarding personal health data, it may not fully capture the spontaneity of real-world information-seeking behaviors. 
\major{In addition, while our naturalistic study design preserved the ecological validity of the search process, it resulted in varying interaction depths across participants. 
For instance, behaviors in the Google condition ranged from ``zero-click'' searches to deep navigation of links to external websites, while ChatGPT interactions varied from single-turn queries to multi-turn dialogues. We acknowledge that the quantity and depth of interaction could have behaved as hidden covariates that influence trust formation. 
Future research should consider longitude and controlled designs (e.g., fixing the number of turns or clicks) to isolate or monitor the specific impact of interaction volume on trust with long-term naturalistic search behaviors.}


Third, our study focused on user perception rather than the factual accuracy of the LLM responses. 
We did not evaluate whether trust levels correlated with the actual accuracy of the information provided. 
As LLMs are prone to ``hallucinations''~\cite{Hallucin28:online}, future work should examine the relationship between response accuracy, hallucination rates, and user trust, particularly as newer tools with web-searching capabilities aim to mitigate these reliability issues.
Additionally, we used OpenAI's native speech recognition to standardize verbal interactions for speech-based and embodied interfaces. While not fully accurate, participants resolved minor errors via rephrasing, and no concerns were raised during our interviews, suggesting minimal impact on the results. Future studies may benefit from integrating more advanced recognition models.


Lastly, \major{the influence of varying demographics, like age and education level, was not fully considered. 
Our samples were recruited primarily from Europe and consisted largely of students and digitally literate users, which may limit the generalizability of the findings to other demographic groups or to users with lower digital literacy.}
Although these factors could potentially influence trust in online information, addressing these variables was beyond the scope of our current work. 
Nevertheless, it provides an opportunity for subsequent research to further explore how different demographic groups perceive and trust online health information differently, thus broadening the understanding of online health information trust across diverse user groups.

With the continuous and rapid advancements in AI technology, it is vital to continue validating and verifying such findings relating to trust perceptions and how these may change over time with newer AI advancements.




\section{Conclusion}

\major{
In this research, we investigated trust in health information seeking by comparing distinct search agents (Google vs. ChatGPT) and three LLM-powered conversational user interfaces (text, speech, and embodied). 
Study 1 revealed that participants placed higher trust in information from ChatGPT than Google, a preference driven by factors such as prior experience, information presentation style, and interaction mode rather than the specific types of health query. 
Study 2 demonstrated that interface modality further influences trust in disseminated information that text-based CUIs being preferred for their familiarity level and ease of use. 
While speech-based and embodied interfaces offered natural verbal interaction but raised concerns about privacy and authenticity.
Collectively, our findings highlight that trust is influenced by both the source agents and dissemination interfaces. 
These insights offer a foundation for developing future health AI tools that are not only reliable and informative but also trusted and tailored to diverse user needs.
}


\newpage


\section*{Disclosure Statement}
The authors declare no conflicts of interest related to this study.





\bibliographystyle{ACM-Reference-Format}

\bibliography{main/trust-interface/ref_trust_theory,main/trust-interface/ref_trust_info,main/trust-interface/ref_trust_web,main/trust-interface/ref_trust_llm,main/trust-interface/ref_trust_cui,main/trust-interface/ref_new}


\end{document}